\documentclass[%
reprint,
superscriptaddress,
 amsmath,amssymb,
 aps,
floatfix,
]{revtex4-2}

\usepackage{graphicx}% Include figure files
\usepackage{dcolumn}% Align table columns on decimal point
\usepackage{bm}% bold math

\usepackage{xparse}
\usepackage{xcolor}

\usepackage{algorithm}
\usepackage{algpseudocode}
\makeatletter
\edef\ftype@algorithm{\the\ftype@algorithm}
\makeatother
\usepackage{hyperref}% add hypertext capabilities (must be loaded after algorithm)
\begin{document}

\title{Remote Entanglement in Lattice Surgery: To Distill, or Not to Distill}% Force line breaks with \\

\author{Sitong Liu}
\email[Corresponding author: ] {sitong.liu@duke.edu}
\affiliation{Duke Quantum Center, Duke University, Durham, NC 27701, USA}
\affiliation{Department of Electrical and Computer Engineering, Duke University, Durham, NC 27708, USA}
\affiliation{Lawrence Berkeley National Laboratory, Berkeley, CA 94720, USA}

\author{John Stack}

\affiliation{Department of Computer Science, North Carolina State University, Raleigh, NC 27606, USA}
\affiliation{Lawrence Berkeley National Laboratory, Berkeley, CA 94720, USA}

\author{Ke Sun}
\affiliation{University of California, Berkeley, CA 94720, USA}
\affiliation{Lawrence Berkeley National Laboratory, Berkeley, CA 94720, USA}

\author{Roel Van Beeumen}
\affiliation{Applied Mathematics and Computational Research Division, Lawrence Berkeley National Laboratory, Berkeley, CA 94720, USA}

\author{Inder Monga}
\affiliation{Lawrence Berkeley National Laboratory, Berkeley, CA 94720, USA}

\author{Katherine Klymko}
\affiliation{National Energy Research Scientific Computing Center, Lawrence Berkeley National Laboratory, Berkeley, CA 94720, USA}

\author{Kenneth R. Brown}
\affiliation{Duke Quantum Center, Duke University, Durham, NC 27701, USA}
\affiliation{Department of Electrical and Computer Engineering, Duke University, Durham, NC 27708, USA}
\affiliation{Department of Physics, Duke University, Durham, NC 27708, USA}
\affiliation{Department of Chemistry, Duke University, Durham, NC 27708, USA}

\author{Erhan Saglamyurek}
\email[Corresponding author: ]{esaglamyurek@lbl.gov}
\affiliation{Lawrence Berkeley National Laboratory, Berkeley, CA 94720, USA}
\affiliation{University of California, Berkeley, CA 94720, USA}

\date{\today}

\begin{abstract}
Distributed quantum computing can potentially address the scalability challenge by networking processors through photon-mediated remote entanglement. Prior approaches assumed that remote Bell pairs require distillation before use, incurring substantial overhead,  to achieve sufficiently high fidelity. However, recent results show that lattice-surgery operations at logical qubit boundaries tolerate significantly higher error rates than previously assumed. We quantify the resource trade-offs between distillation overhead and surface-code distance requirements under realistic constraints including probabilistic entanglement generation and memory decoherence. We identify the fidelity crossover point separating the two regimes. Below this threshold, the distillation strategy dominates, reducing resource overhead by up to two orders of magnitude. Above it, no-distillation becomes the more efficient choice, reducing resource overhead by more than half. We briefly describe the application of these methods to ion-trap and neutral-atom platforms. These results provide joint design guidelines for optimizing photonic interconnects and fault-tolerant architectures in distributed quantum computing.

\end{abstract}

\maketitle

%\tableofcontents

\section{\label{sec:intro}Introduction}

As quantum processors scale, building ever-larger single-chip devices becomes increasingly difficult due to practical hardware limits such as chip area, control crosstalk, and thermal and optical routing constraints. Modular or distributed quantum computing (DQC) addresses these challenges by networking multiple medium-scale processors through photonic entanglement links. In such architectures, fault tolerance is typically provided by quantum error correction (QEC), while nonlocal logical operations are mediated by shared Bell pairs used for teleportation~\cite{jacinto2025networkrequirementsdistributedquantum,noisylink2024,shalby2025optimizednoiseresilientsurfacecode,haug2025latticesurgerybellmeasurements,stack2025assessingteleportationlogicalqubits}.

The achievable performance in DQC is jointly constrained by local noise within each module and by the fidelity, availability, and consumption of entanglement distributed between modules. Moreover, conventional fault-tolerant distributed approaches typically assume that raw Bell pairs must undergo entanglement distillation \cite{Dr2002EntanglementPF} before use, as the Bell-pair fidelity is often below the error-correction threshold~\cite{Pathumsoot:2024lbu,leone2024resourceoverheadsattainablerates, de_Bone_2024}, resulting in substantial overhead. Dedicated distillation factories and ancillary communication qubits can consume a substantial fraction of a module's physical resources, motivating the search for regimes in which distillation can be bypassed.

\begin{figure*}[ht]
  \centering
  \includegraphics[width=\textwidth]{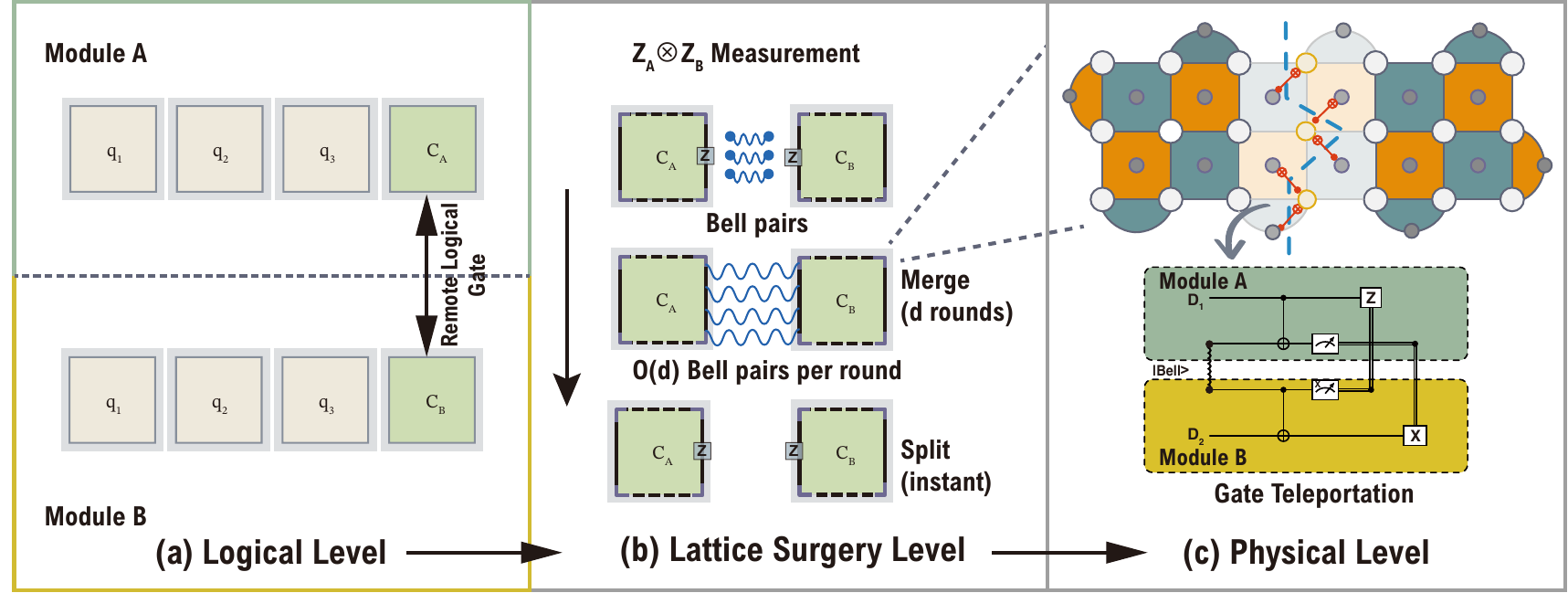}%

  \caption{\textbf{Remote lattice surgery for distributed architecture.}
  (a) Logical level: Modules A and B are physically separated and perform remote operations on encoded logical qubits through communication qubits (green) via Pauli product measurements.
  (b) Lattice surgery level: $Z \otimes Z$ measurements between logical qubits C$_A$ and C$_B$ are implemented via merge-split of the logical qubit boundaries. The merge operation requires $d$ syndrome measurement rounds, where $d$ is the surface code distance, consuming $\mathcal{O}(d)$ Bell pairs per round. The split operation is instantaneous as it coincides with the standard syndrome measurement cycle. (c) Physical level: Top illustrates the merge operation between two surface code patches encoding logical qubits q$_1$ (left) and q$_2$ (right). Data qubits are shown as white circles and stabilizer plaquettes as colored regions. The light yellow circles and faded areas represent ancilla qubits and additional stabilizers generated during merging. Blue dashed lines indicate the physical module boundaries. Bottom shows the gate teleportation circuit with Bell pair. The Bell pairs connect the interface regions to enable the remote CNOT operations (marked in red in the top panel).}

  \label{fig:remote_surgery}
\end{figure*}

A recent insight substantially changes this picture. QEC simulations show that errors at surface-code patch interfaces have significantly higher fault-tolerance thresholds during lattice surgery. These thresholds are approximately one order of magnitude higher than errors within individual code patches. For example, using a circuit-level noise model, studies show a Bell-pair error threshold of $15.3\%$ for remote gate teleportation when local physical noise is $0.1\%$~\cite{sunami2025entanglementboostinglowvolumelogical}. This asymmetric tolerance means moderate-fidelity Bell pairs may suffice for remote lattice surgery without prior purification, forming the basis for the distillation-free approach we examine here.

However, whether this regime is practically accessible depends on implementation constraints. Prior studies of modular and distributed architectures often model inter-processor links as effective channels with fixed rate and error parameters. This simple assumption hides a fundamental conflict that entanglement between modules is generated probabilistically and must be consumed within well-defined operational windows set by QEC cycles. Under these constraints, the feasibility of distributed surface-code operations depends sensitively on how many raw Bell pairs are required per remote lattice surgery measurement, how their fidelity evolves during buffering and use, and how many logical patches and communication resources can be hosted within each processor, etc. These constraints become particularly strict in regimes that seek to avoid entanglement distillation, where raw Bell pairs are consumed directly without intermediate purification.

In this work, we establish the intrinsic crossover boundary that determines where distillation-free and distillation-assisted architectures are respectively favorable. Strict physical trade-offs among Bell-pair consumption, Bell-pair fidelity, and local processor size dictate this boundary. By highlighting the key trade-offs in the co-design of photonic interconnects and fault-tolerant logical layouts, these results provide quantitative guidance for designing modular surface-code architectures.

\section{Results}
\subsection{Distillation optimality criterion under static fidelity \label{sec: static}}

\textbf{To distill or not to distill, that is the question.}

\begin{figure}[htbp]
  \centering
  \includegraphics[width=\columnwidth]{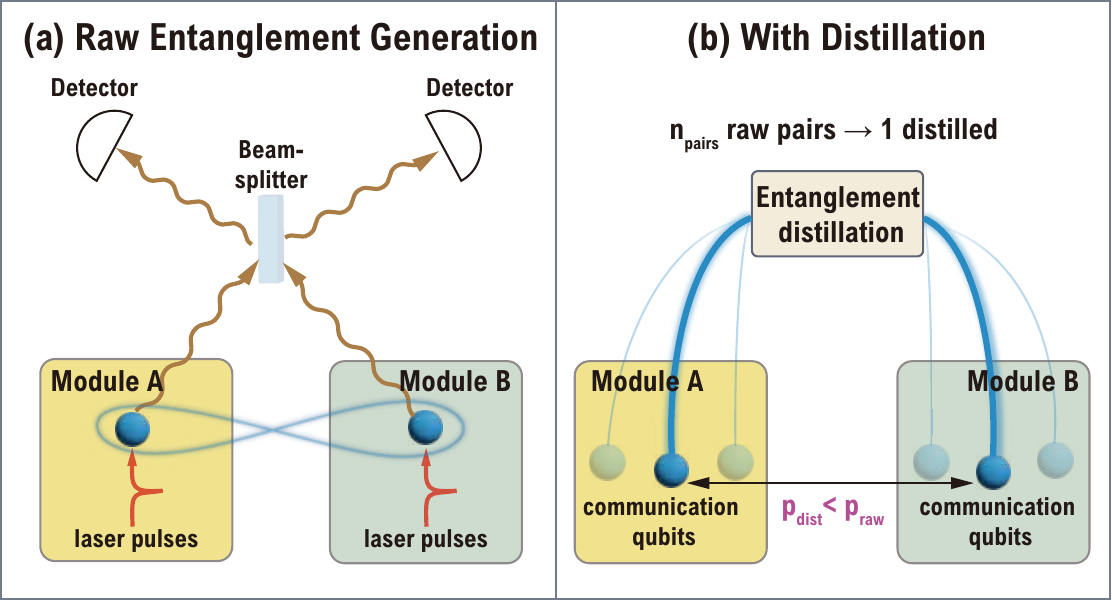}
  \caption{\textbf{Comparison of remote entanglement utilization schemes.}
  (a) Direct use of raw Bell pairs. Heralded entanglement is generated
  via optical links between communication qubits in Module~A and~B,
  with link error $p_{\mathrm{raw}}$. (b) Distillation-based approach.
  $n_{\mathrm{pairs}}$ raw Bell pairs are consumed by an entanglement distillation
  protocol $D$ to produce a single high-fidelity Bell pair with error
  $p_{\mathrm{eff}} < p_{\mathrm{raw}}$.}
  \label{fig:distillation}
\end{figure}

We now quantify when direct consumption of raw Bell pairs is resource-optimal over distillation, as a function of Bell-pair fidelity and distillation overhead, for a given target logical error rate.

Each round of remote stabilizer measurement in lattice surgery couples $\mathcal{O}(d_s)$
boundary qubits via distributed entanglement~\cite{Litinski2019gameofsurfacecodes,
jacinto2025networkrequirementsdistributedquantum}, consuming
\begin{equation}
    n^{\mathrm{round}} = a d_s - c
    \label{eq:round_pairs}
\end{equation}
Bell pairs, where $(a,c)=(2,1)$ for gate-teleportation (see Fig.~\ref{fig:remote_surgery}(c))~\cite{sunami2025entanglementboostinglowvolumelogical}
and $(a,c)\approx(2\text{--}1,\,0)$ for measurement-teleportation and Bell-measurement
schemes~\cite{jacinto2025networkrequirementsdistributedquantum,haug2025latticesurgerybellmeasurements}.
Since each logical operation requires $d_s$ consecutive syndrome-extraction
rounds~\cite{Litinski2019gameofsurfacecodes}, the total Bell-pair consumption is
\begin{equation}
    N_{\mathrm{QEC}} = d_s \cdot n^{\mathrm{round}} = a d_s^2 - c d_s.
    \label{eq:N_QEC}
\end{equation}

\begin{figure*}[ht]
\centering
\includegraphics[width=0.49\textwidth]{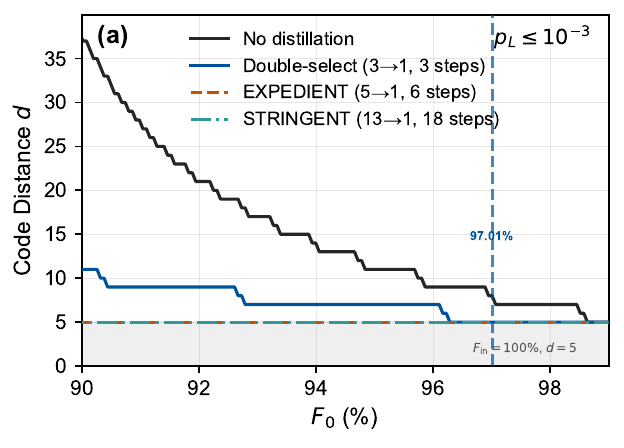}
\hfill
\includegraphics[width=0.49\textwidth]{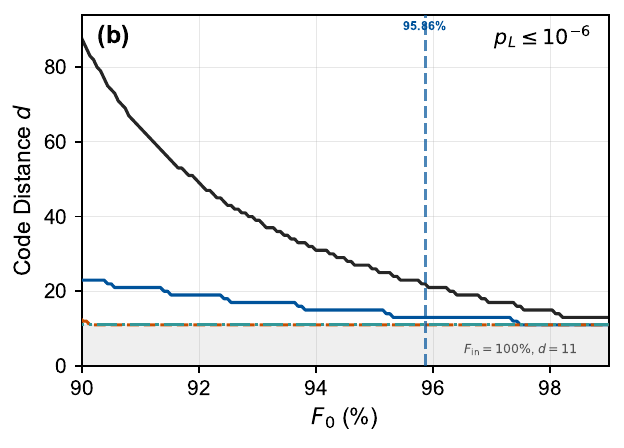}
\\[1em]
\includegraphics[width=0.49\textwidth]{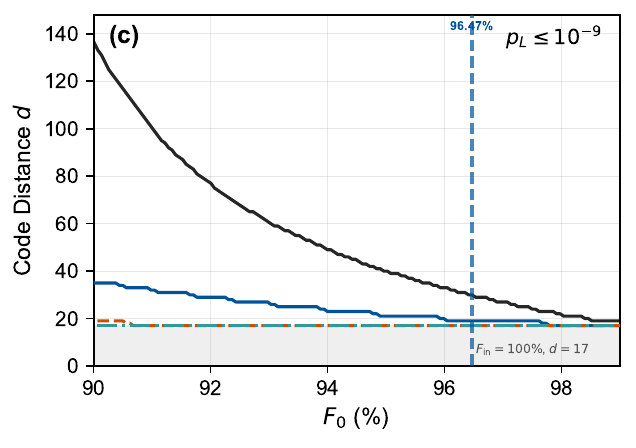}
\hfill
\includegraphics[width=0.49\textwidth]{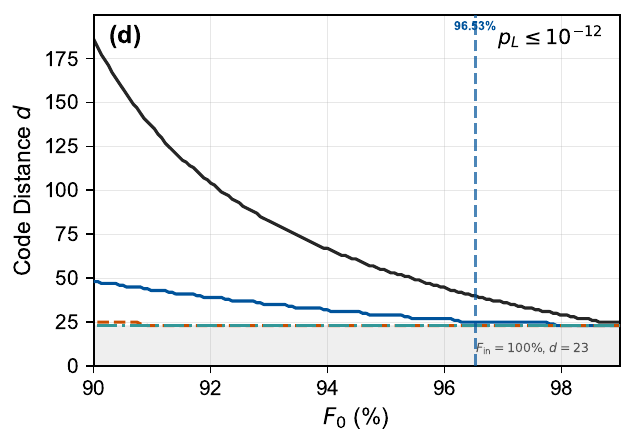}
\caption{\textbf{Required surface-code distance $\bm{d}$ for remote
  lattice surgery versus raw Bell-pair fidelity
  $\bm{F_0}$, at target logical error rates
  (a)~$\bm{p_L \le 10^{-3}}$, (b)~$\bm{10^{-6}}$,
  (c)~$\bm{10^{-9}}$, (d)~$\bm{10^{-12}}$.}
  The no-distillation baseline (black solid) is compared with three
  entanglement distillation protocols from
  Ref.~\cite{Krastanov2019optimized}:
  double-select (blue solid, $3{\to}1$, 3~steps),
  EXPEDIENT (orange dashed, $5{\to}1$, 6~steps),
  and STRINGENT (green dash-dot, $13{\to}1$, 18~steps).
  Grey-shaded regions indicate the minimum achievable distance at
  perfect input fidelity ($F_0 = 100\%$), where only local errors
  contribute.
  The vertical dashed line in each panel marks the fidelity at which
  no distillation first achieves a shorter total operation time
  than every distillation protocol (see Sec.~\ref{subsubsec:time}); its color matches the last
  protocol overtaken (double-select, blue). Distance--fidelity model from
  Eq.~(\ref{eq:bell-logical-error}); error parameters in
  Table~\ref{tab:noise_parameters}.}

\label{fig:distance_four_panels}

\end{figure*}

\begin{figure*}[ht]
\centering
\includegraphics[width=0.49\textwidth]{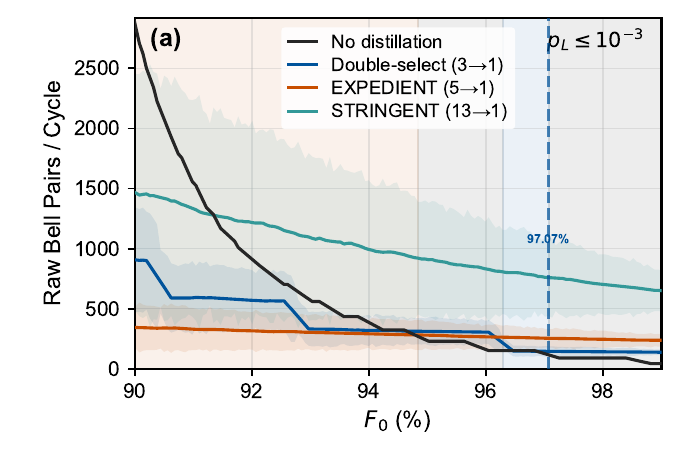}
\hfill
\includegraphics[width=0.49\textwidth]{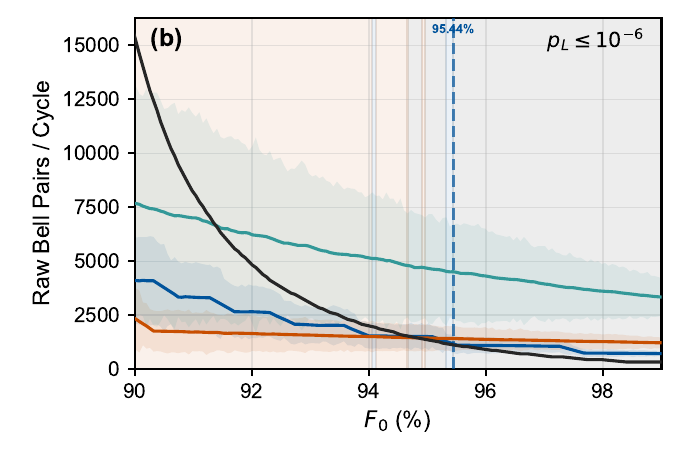}
\\[1em]
\includegraphics[width=0.49\textwidth]{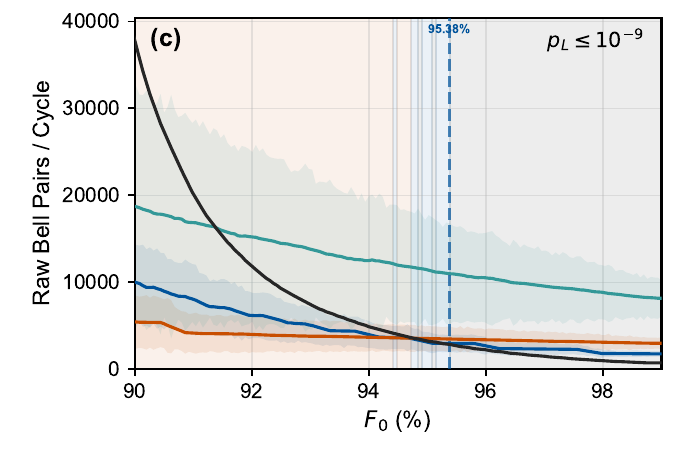}
\hfill
\includegraphics[width=0.49\textwidth]{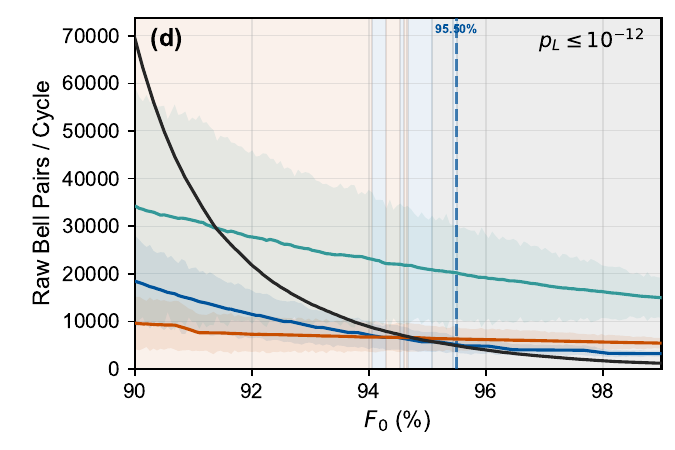}
\caption{\textbf{Raw Bell-pair consumption per QEC cycle between
  two logical qubits via remote lattice surgery versus raw Bell-pair
  fidelity ($\bm{F_0}$), assuming
  no memory decoherence, at target logical error rates
  (a)~$\bm{p_L \le 10^{-3}}$, (b)~$\bm{10^{-6}}$,
  (c)~$\bm{10^{-9}}$, (d)~$\bm{10^{-12}}$.}
  Curves compare the no-distillation case (black) with three
  entanglement distillation protocols from
  Ref.~\cite{Krastanov2019optimized}: double-select $3{\to}1$
  (blue), EXPEDIENT $5{\to}1$ (orange), and STRINGENT $13{\to}1$
  (green).
  Cost quantifies total raw Bell-pair consumption, incorporating
  surface-code distance scaling ($\propto d_s^2$) and distillation
  overhead.
  Colored background regions indicate the resource-optimal strategy
  for different input fidelity regimes.
  The vertical dashed line marks the fidelity above which
  no~distillation is optimal; its colour (blue) indicates the
  last protocol overtaken (double-select in all cases).
  No~distillation is optimal for
  $F_0 \gtrsim 97.07\%$~(a),
  $95.44\%$~(b),
  $95.38\%$~(c), and
  $95.50\%$~(d).
  The greatest no-distillation advantage is a~$68\%$ overhead
  reduction ($F_0 = 98.64\%$, $p_L = 10^{-3}$:
  45 vs 140~pairs/cycle, both $d_s = 5$).
  Shaded bands show Monte Carlo uncertainty from $10^3$~runs at
  150~fidelity values spanning $F_0 \in [90.0\%, 99.0\%]$.
  Distillation parameters are taken from Ref.~\cite{Krastanov2019optimized}; distance--fidelity model is from Eq.~(\ref{eq:bell-logical-error});
  error rates are in Table~\ref{tab:noise_parameters}.}
\label{fig:cost_four_panels}
\end{figure*}

Entanglement distillation maps $n_{\mathrm{pairs}}$ raw Bell pairs 
with error rate $p_{\mathrm{raw}}$ to a single pair with reduced 
error rate $p_{\mathrm{eff}} = f_{\mathcal{D}}(p_{\mathrm{raw}})$, 
succeeding with probability $p_{\mathrm{succ}}$; 
the resulting distance ratio 
$\rho \equiv d_s^*(p_{\mathrm{eff}})/d_s^*(p_{\mathrm{raw}}) \in (0,1]$, where $d_s^{*}(p_{\mathrm{Bell}})$ denotes the minimum code distance as a function of Bell-pair error rate $p_{\mathrm{Bell}}$,
quantifies the code-distance reduction.

We compare the total Bell-pair consumption of using raw Bell pairs versus entanglement distillation per complete lattice-surgery operation (spanning $d_s$ syndrome rounds):

\begin{equation}
\begin{split}
C_{\mathrm{raw}}^{\mathrm{cycle}}
&= d_s^{*}(p_{\mathrm{raw}}) \cdot \left(a d_s^{*}(p_{\mathrm{raw}}) - c\right), \\
C_{\mathrm{dist}}^{\mathrm{cycle}}
&= \frac{n_{\mathrm{pairs}}}{p_{\mathrm{succ}}} \cdot d_s^{*}(p_{\mathrm{eff}}) \cdot \left(a d_s^{*}(p_{\mathrm{eff}}) - c\right),
\end{split}
\label{eq:costs_cycle}
\end{equation}
substituting $d_s^{*}(p_{\mathrm{eff}}) = \rho d_s^{*}(p_{\mathrm{raw}})$, the cost ratio becomes

\begin{equation}
\frac{C_{\mathrm{dist}}^{\mathrm{cycle}}}{C_{\mathrm{raw}}^{\mathrm{cycle}}}
= \frac{n_{\mathrm{pairs}}}{p_{\mathrm{succ}}} \cdot \frac{d_s^{*}(p_{\mathrm{eff}}) \left[a d_s^{*}(p_{\mathrm{eff}}) - c\right]}{d_s^{*}(p_{\mathrm{raw}}) \left[a d_s^{*}(p_{\mathrm{raw}}) - c\right]}
\approx \frac{n_{\mathrm{pairs}}}{p_{\mathrm{succ}}} \cdot \rho^2,
\label{eq:cost_ratio}
\end{equation}
where the approximation neglects $c$ relative to $a d_s^{*}(p_{\mathrm{raw}})$ and is accurate to within 10\% for typical parameters ($a=2, c=1, d_s \geq 7$). Distillation becomes resource-optimal when this ratio falls below unity, i.e., when the quadratic distance reduction ($\rho^2$) more than compensates for the consumption overhead ($n_{\mathrm{pairs}}/p_{\mathrm{succ}}$).

\textbf{Direct consumption of raw Bell pairs is resource-optimal when the overhead from entanglement distillation exceeds the benefit from code-distance reduction}:

\begin{equation}
\boxed{\frac{C_{\mathrm{dist}}^{\mathrm{cycle}}}{C_{\mathrm{raw}}^{\mathrm{cycle}}}
= \frac{n_{\mathrm{pairs}}}{p_{\mathrm{succ}}} \cdot \rho^2
\gtrsim 1,}
\label{eq:nodist_optimal}
\end{equation}

Distillation is favored when the quadratic distance gain $\rho^2$
outweighs the overhead factor $n_{\mathrm{pairs}}/p_{\mathrm{succ}}$,
i.e., at low fidelities where $\rho \ll 1$; at high fidelities,
$\rho \to 1$ and the overhead is never justified.

We analyze, in this work, representative protocols with $n_{\mathrm{pairs}} \in \{1, 3, 5, 13\}$,
corresponding to no distillation and the following entanglement distillation
protocols: double selection~\cite{Krastanov2019optimized, Fujii2008EntanglementPW},
EXPEDIENT, and STRINGENT~\cite{Krastanov2019optimized, strigent,
Fujii2008EntanglementPW}. These three protocols outperformed other well-known
candidates, including BBPSSW~\cite{PhysRevLett.76.722} and
DEJPMS~\cite{PhysRevLett.77.2818}, across our parameter regime, and
genetic-algorithm searches~\cite{Krastanov2019optimized} have not identified a
protocol that uniformly surpasses them at all input fidelities. Figure 4 demonstrates this trend across four target logical error rates: the crossover fidelity is
remarkably stable at $F_0 \approx 95$--$97\%$ over
$p_L \in [10^{-3}, 10^{-12}]$, indicating that the optimal
strategy is governed primarily by Bell-pair quality rather than how tight the target error rate is. 

\subsection{\label{subsec:decay}Distillation trade-off under time-dependent decoherence}

The preceding analysis assumed Bell pairs are immediately available upon request without decoherence during storage. Under realistic hardware constraints, however, entanglement generation proceeds at a finite rate determined by the attempt frequency and heralding success probability, while accumulated pairs decohere during waiting periods. This section estimates how these temporal effects modify the distillation trade-off.

\subsubsection{Time overhead of distillation}
\label{subsubsec:time}

We compare execution time using circuit depth as a proxy, since 
two-qubit gate time and measurement time dominate the total circuit 
duration (see Table~\ref{tab:timing}). Here, execution time refers 
purely to circuit runtime and does not include Bell-pair generation 
latency or buffering time. Let $\tau_{\mathrm{SE}}$ denote the 
execution time of one syndrome extraction round. We denote the 
protocol-dependent per-round duration and Bell-pair consumption by 
$T_{\mathrm{round}}$ and $C_{\mathrm{round}}$, respectively:
\begin{align}
T^{\mathrm{round}}_{\mathrm{raw}} &= \tau_{\mathrm{SE}}, 
&C^{\mathrm{round}}_{\mathrm{raw}} &= n^{\mathrm{round}} = ad_s - c, 
\label{eq:round_raw} \\
T^{\mathrm{round}}_{\mathrm{dist}} &= \tau_{\mathrm{D}} + \tau_{\mathrm{SE}}, 
&C^{\mathrm{round}}_{\mathrm{dist}} &= \frac{n_{\mathrm{pairs}}}{p_{\mathrm{succ}}} \cdot n^{\mathrm{round}},
\label{eq:round_dist}
\end{align}
where we assume all purification circuits for one syndrome round 
run in parallel, completing before the syndrome extraction begins, 
so only one distillation depth $\tau_{\mathrm{D}}$ is added per round. 
A full QEC cycle of $d_s$ rounds therefore costs
\begin{equation}
\label{eq:T_dist}
T^{\mathrm{cycle}}_{\mathrm{raw}} = \tau_{\mathrm{SE}}\,d_{\mathrm{raw}}, 
\qquad
T^{\mathrm{cycle}}_{\mathrm{dist}} = (\tau_{\mathrm{D}} + \tau_{\mathrm{SE}})\,d_{\mathrm{dist}},
\end{equation}
where $d_{\mathrm{raw}} \equiv d_s^{*}(p_{\mathrm{raw}})$ and 
$d_{\mathrm{dist}} \equiv d_s^{*}(p_{\mathrm{eff}})$.
Distillation reduces total execution time when $T^{\mathrm{cycle}}_{\mathrm{dist}} < T^{\mathrm{cycle}}_{\mathrm{raw}}$,
i.e., $(\tau_{\mathrm{D}} + \tau_{\mathrm{SE}})\,d_{\mathrm{dist}} <
\tau_{\mathrm{SE}}\,d_{\mathrm{raw}}$, which rearranges to
\begin{equation}
\frac{d_{\mathrm{dist}}}{d_{\mathrm{raw}}}
\;<\; \frac{1}{1 + \tau_{\mathrm{D}}/\tau_{\mathrm{SE}}}.
\label{eq:time_threshold}
\end{equation}

For STRINGENT, which has the largest circuit depth among the
protocols considered ($\approx 18$ timesteps $\approx 3.6\,\tau_{\mathrm{SE}}$),
this threshold corresponds to $d_{\mathrm{dist}}/d_{\mathrm{raw}} \lesssim 0.22$.
In the low-fidelity regime where $d_{\mathrm{dist}} \ll d_{\mathrm{raw}}$,
Eq.~\eqref{eq:time_threshold} is easily satisfied and distillation
reduces $T$ substantially.
As raw fidelity increases and $d_{\mathrm{dist}}/d_{\mathrm{raw}} \to 1$,
the condition is violated and distillation becomes slower than direct use.
Beyond execution time, Bell-pair consumption $N_{\mathrm{QEC}}$
follows a steeper $d^2$ scaling (Eq.~\eqref{eq:nodist_optimal}),
with $\rho^2$ varying more rapidly than $\rho$ as $\rho \to 1$;
the relative position of the two crossovers is set by the
prefactors $(1+\tau_{\mathrm{D}}/\tau_{\mathrm{SE}})$ and
$(n_{\mathrm{pairs}}/p_{\mathrm{succ}})$ respectively.
Fig.~\ref{fig:distance_four_panels} shows the fidelity above which the
no-distillation protocol achieves shorter execution time than
every distillation protocol (Eq.~\eqref{eq:time_threshold});
Fig.~\ref{fig:cost_four_panels} shows the corresponding crossover 
in raw Bell-pair consumption per QEC cycle, where the 
no-distillation cost (Eq.~\eqref{eq:nodist_optimal}) is compared against that of the best-performing distillation protocol at each fidelity point.

\subsubsection{Rate-Limited operating regimes\label{subsec:decay_choice}}

The operating regime is determined by comparing the expected number 
of Bell pairs generated per round, $n_{\mathrm{gen}} \equiv \lambda T^{\mathrm{round}}$, 
where $\lambda$ (pairs\,s$^{-1}$) is the heralded generation rate 
(see Sec. \ref{subsec:entanglement_generation}), 
to the per-round consumption $C^{\mathrm{round}}$.
Since photon-mediated entanglement attempts succeed independently, 
Bell pairs arrive as a Poisson process with rate $\lambda$; 
the number generated in one round of duration $T^{\mathrm{round}}$ 
is therefore $N \sim \mathrm{Poisson}(\lambda T^{\mathrm{round}})$.

\begin{figure*}[ht]
  \centering \includegraphics[width=\textwidth]{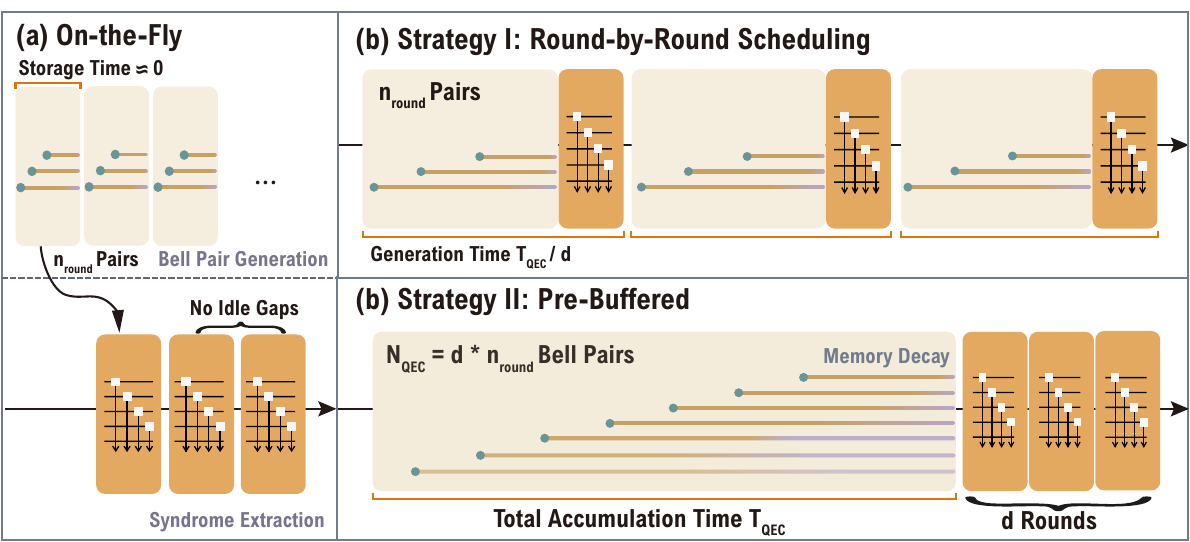}
\caption{%
\textbf{Bell-pair scheduling schemes for remote lattice surgery
above and below the on-the-fly threshold
\bm{$\lambda_{\mathrm{th}}$}
(Eq.~\eqref{eq:on_the_fly_condition}).}
(a)~On-the-fly regime ($\lambda \geq \lambda_{\mathrm{th}}$):
the generation rate is sufficient to supply each syndrome round
on demand, so Bell pairs are consumed immediately with
worst-case storage time bounded by $T^{\mathrm{round}}$
(Eq.~\eqref{eq:fidelity_decay_bound}).
(b)~When generation cannot keep pace ($\lambda < \lambda_{\mathrm{th}}$),
two strategies arise.
\emph{Strategy~1} (round-by-round):
collecting
$n^{\mathrm{round}}$ pairs per round requires a finite
accumulation window; pairs generated early in the window
must wait in memory until the round begins, degrading their
fidelity (Eq.~\eqref{eq:fidelity_S1}), and data qubits
likewise accumulate idle error during the wait
(Eq.~\eqref{eq:idle_error}).
\emph{Strategy~2} (pre-buffered):
all $N_{\mathrm{QEC}}$ pairs are collected before any
syndrome extraction begins; the $\tilde{d}_s$ rounds then
proceed back to back, eliminating data-qubit idle error but
exposing the earliest pairs to decay over the full
accumulation time $T_{\mathrm{QEC}}$
(Eq.~\eqref{eq:fidelity_S2}).
In both strategies the required code distance
$\tilde{d}_s$ and the per-round consumption
$n^{\mathrm{round}}$ are mutually dependent, requiring
the self-consistent iteration of
Eq.~\eqref{eq:self_consistent_ds}
(Algorithm~\ref{alg:self_consistent}).}
\label{fig:scheduling}
\end{figure*}

\paragraph{On-the-fly operation\label{subsubsec:onthefly}} 
In the on-the-fly regime, entanglement generation and consumption
run in parallel, with no idling between syndrome measurement rounds. 

Applying the normal approximation 
$N \approx \mathcal{N}(\lambda T^{\mathrm{round}},\,\lambda T^{\mathrm{round}})$, where mean and variance both equal
the Poisson rate $\lambda T^{\mathrm{round}}$.
When entanglement generation keeps pace with consumption within each
round, the \textit{on-the-fly} condition (OTF) reads
\begin{equation}
\lambda \, T^{\mathrm{round}}
  - \Phi^{-1}(0.99)\sqrt{\lambda\, T^{\mathrm{round}}}
  \;\geq\;
  C^{\mathrm{round}}.
\label{eq:on_the_fly_condition}
\end{equation}

In the on-the-fly regime, the worst-case storage time is
$t_{\max} \lesssim T^\mathrm{round}$. Under the exponential decoherence model
(Sec.~\ref{subsub:decay}), the stored Bell-pair fidelity $F_{\mathrm{stored}}$
is therefore bounded by
\begin{equation}
F_{\mathrm{stored}}
  \geq F_0\,e^{-T^\mathrm{round}/\tau_{\mathrm{coh}}}
  \approx F_0\!\left(1
        - \frac{T^\mathrm{round}}{\tau_{\mathrm{coh}}}\right).
\label{eq:fidelity_decay_bound}
\end{equation}
The related question of when to distill, including whether to process pairs as they arrive or to batch them, is studied in
Ref.~\cite{yakar2025advantagesglobalentanglementdistillationpolicies}. Here we assume all raw pairs for one round are available at the start of the round. The raw input pairs for distillation are stored for a longer time,
experiencing up to $T^{round}_{\mathrm{dist}}/T^{round}_{\mathrm{raw}}$ times the
exposure duration compared to pairs consumed directly without
distillation (Eqs.~\eqref{eq:round_raw}--\eqref{eq:round_dist}). For the protocols considered in this work,
$T^{round}_{\mathrm{dist}} \lesssim 5\,T^{round}_{\mathrm{raw}}$. Since the
storage time is short compared with the coherence time, the exponential
fidelity decay is approximately linear, so a raw pair entering distillation
acquires a storage-induced error increment $\delta p_{\mathrm{raw}}$ that scales
with this exposure ratio. Double-selection and higher distillation protocols
remove all first-order input
errors~\cite{Fujii2008EntanglementPW,Krastanov2019optimized}, so the purified
output depends on the raw error only at second order. The increment
$\delta p_{\mathrm{raw}}$ therefore raises the output by only
$\Delta p_{\mathrm{eff}}=O(p_{\mathrm{raw}}\,\delta p_{\mathrm{raw}})$, with a
smaller cross-term contribution $O(p_{\mathrm{local}}\,\delta p_{\mathrm{raw}})$,
i.e.\ suppressed by a factor of order the raw Bell-pair error
$p_{\mathrm{raw}}=1-F_0$. With the crossover at
$p_{\mathrm{raw}}>p_{\mathrm{local}}$ and $p_{\mathrm{raw}}\lesssim 10\%$ in the
regime of interest, this factor is small, so storage perturbs the purified
output only weakly despite the longer exposure. The directly consumed pairs are
stored only for the shorter $T^{round}_{\mathrm{raw}}$ and likewise acquire only
a small storage shift in this regime, so the crossover is only weakly affected
by storage decoherence.

\paragraph{No-expire regime}

\begin{figure*}[ht]
\centering
\includegraphics[width=0.49\textwidth]{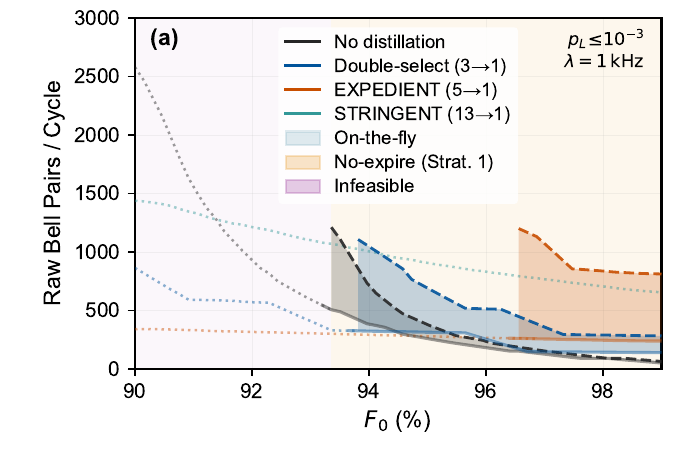}\hfill
\includegraphics[width=0.49\textwidth]{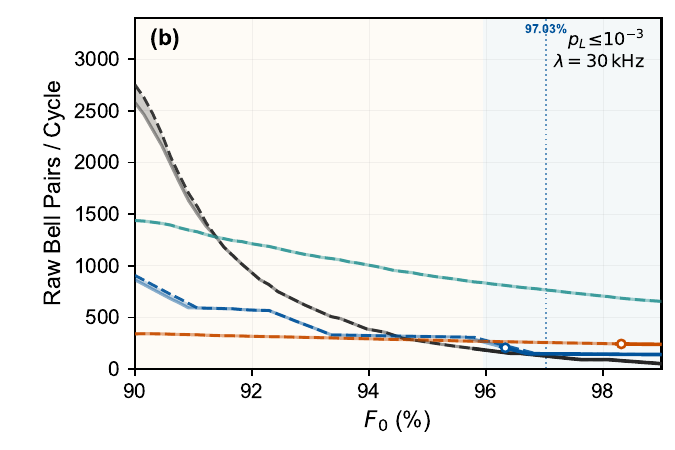}\\[4pt]
\includegraphics[width=0.49\textwidth]{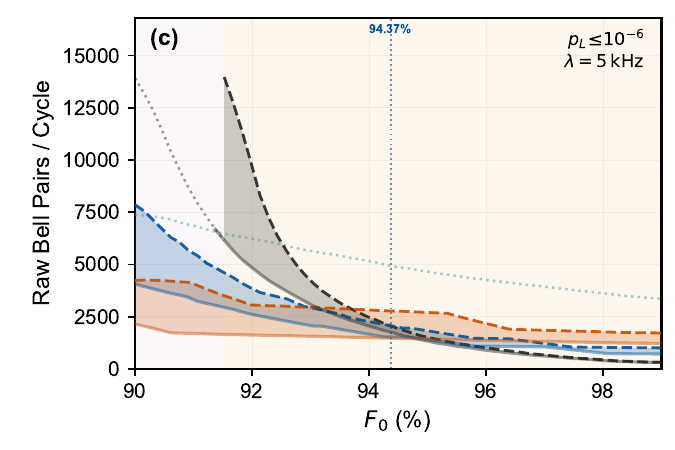}\hfill
\includegraphics[width=0.49\textwidth]{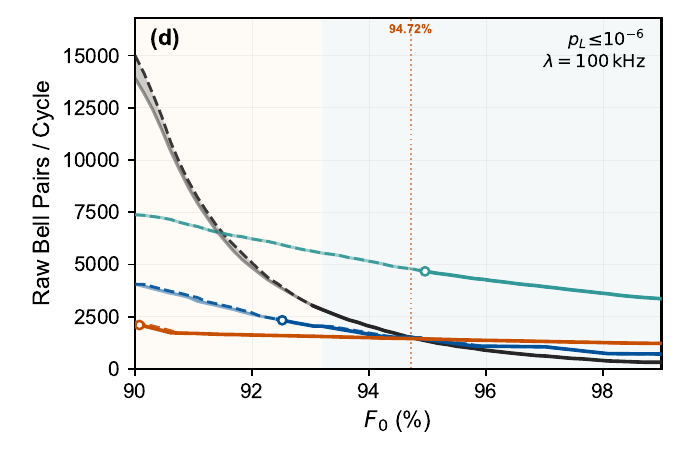}
\caption{\textbf{Operating regimes and resource cost at four
representative $(\boldsymbol{p_L},\,\boldsymbol{\lambda})$
settings.}
  Fixed parameters:
  $\tau_{\mathrm{SE}} = 1\,$ms,
  $\tau_{\mathrm{coh}} = 10\,$s,
  $\tau_{\mathrm{dep}} = 50\,$s
  ($\mu = 5$).
  Background shading marks the no-distillation regime boundaries:
  \emph{on-the-fly} (green, Eq.~\eqref{eq:on_the_fly_condition}),
  \emph{no-expire} (yellow, Eq.~\eqref{eq:self_consistent_ds}),
  and \emph{infeasible} (pink). Cost curves are shown for all four protocols: solid lines show the static (no-decay) cost: full opacity in the
  on-the-fly zone, reduced opacity in the no-expire zone, and dotted
  in infeasible regions;
  an open circle marks the OTF\,/\,no-expire transition for each
  distillation protocol.
  Dashed lines show the converged no-expire cost, in which both
  Bell-pair decay and data-qubit idle depolarization
  (Eq.~\eqref{eq:idle_error}) feed back into~$\tilde{d}_s$;
  the shaded fill between solid and dashed curves visualises
  the resulting cost overhead.
  Panel~(a): at~$\lambda = 1\,\mathrm{kHz}$, no-distillation is the
  cheapest protocol across the entire feasible~$F_0$ range.
  Panels~(a)\,--\,(b): raising~$\lambda$
  from~$1$ to~$30\,\mathrm{kHz}$ eliminates the infeasible
  zone and opens the OTF regime at moderate~$p_L$;
  panels~(c)\,--\,(d): at the stricter target
  $p_L\!\leq\!10^{-6}$, raising~$\lambda$ from~$5$
  to~$100\,\mathrm{kHz}$ likewise removes infeasibility
  and enables on-the-fly operation.}
  \label{fig:regime-classification}
\end{figure*}

When $\lambda < n^{\mathrm{round}}/T^{\mathrm{round}}$ but
the link efficiency $\eta_{\mathrm{link}} \equiv \lambda\,\tau_{\mathrm{coh}}$
is large enough that the earliest pair has  not decayed below the error correction threshold $F_{\mathrm{th}}$ by the time all $N_{\mathrm{QEC}}$ pairs are
collected, the system operates in the no-expire regime.
The full batch arrives slower than one syndrome cycle, so
the surface-code patch idles while waiting for collection
to complete.

Two scheduling strategies can accommodate this regime
(Fig.~\ref{fig:scheduling}(b)): \emph{round-by-round}, which
feeds pairs to the code as each syndrome round becomes ready,
and \emph{pre-buffered}, which accumulates all $N_{\mathrm{QEC}}$ pairs
before starting extraction. A detailed comparison in Sec.~\ref{app:scheduling_comparison} shows
that Strategy~1 outperforms Strategy~2 in the regimes of interest,
as round-by-round collection limits storage to at most one round
whereas pre-buffering forces early pairs to wait the full
accumulation time. We therefore adopt Strategy~1 throughout.

Under Strategy~1, the earliest pair in each round decays to
\begin{equation}
  F_{\mathrm{stored}}
  \;\approx\;
  F_0\,e^{-C^{\mathrm{round}}/\eta_{\mathrm{link}}},
  \label{eq:fidelity_S1}
\end{equation}

This fidelity degradation increases the required code distance
through a self-consistency loop: a larger $d_s$ demands more Bell
pairs per round, lengthening collection and causing further decay. To quantify how finite generation rate and memory decoherence
reshape the distillation trade-off, we solve the self-consistent
distance equation (Eq.~\eqref{eq:self_consistent_ds}) over a range of fidelities
$F_0 = 90$--$99\,\%$ at several representative
$(p_L,\,\lambda)$ settings
(Fig.~\ref{fig:regime-classification}).
Two decoherence time scales govern the penalty:
$\tau_{\mathrm{coh}}$, the characteristic time for a stored
Bell pair to depolarise toward the maximally mixed state, and
$\tau_{\mathrm{dep}} = \mu\,\tau_{\mathrm{coh}}$, the
characteristic time for an idling data qubit to accumulate
errors between syndrome rounds. The ratio
$\mu \equiv \tau_{\mathrm{dep}}/\tau_{\mathrm{coh}}$ is
platform-dependent. For two nodes of the same platform, a
stored Bell pair spans two halves that decohere independently
at the same rate, so for Bell-pair and data qubits of the same
type it depolarises twice as fast as a single data qubit,
giving a baseline $\mu = 2$. Platform-specific effects shift
this baseline, most directly the storage of the Bell pair on a
different type of memory qubit, which moves $\mu$ depending on
the relative coherence of the memory and data qubits. We adopt $\mu=5$ as the baseline for the cost analysis (Fig.~\ref{fig:regime-classification}).

At the generation rates considered
($\lambda = 5$--$100\,\mathrm{kHz}$), this gives link efficiencies
$\eta_{\mathrm{link}} = 5\times 10^{4}$--$10^{6}$,
spanning the no-expire through on-the-fly regimes.
Because all decoherence enters through
$\eta_{\mathrm{link}}$, the results scale directly with memory
lifetime: longer $\tau_{\mathrm{coh}}$ widens the no-expire
window and shrinks the idle penalty, while shorter
$\tau_{\mathrm{coh}}$ pushes the system towards infeasibility.
The ratio $\mu$ enters the model only through the data-qubit idle error $p_{\mathrm{idle}}$
(Eq.~\eqref{eq:idle_error}); increasing it suppresses the gap
between the static and no-expire cost curves, whereas
$\mu \to 1$ amplifies the penalty and narrows the feasible
region.

Figure~\ref{fig:regime-classification} compares the static (solid)
and converged Strategy~1 (dashed) costs across four
$(\lambda,\,p_L)$ settings.
In the OTF zone (panels~b,\,d) the cost tracks the static
prediction; the crossover analysis of
Sec.~\ref{sec: static} applies directly.
In the no-expire regime the self-consistency
loop~\eqref{eq:self_consistent_ds} inflates~$\tilde{d}_s$,
opening the shaded gap between the two curves.
As $C^{\mathrm{round}}_{\mathrm{dist}} = (n_{\mathrm{pairs}}/p_{\mathrm{succ}})
\times n^{\mathrm{round}}(\tilde{d}_s)$,
the feedback gain is amplified by
$n_{\mathrm{pairs}}/p_{\mathrm{succ}}$;
raw Bell pairs ($C^{\mathrm{round}}_{\mathrm{raw}}$) have the
weakest feedback, while STRINGENT ($13\!:\!1$) has the strongest
and is the first to become infeasible. Because temporal decoherence penalises distillation more heavily
than no-distillation,
the crossover fidelity generically shifts downward
(toward lower~$F_0$) or disappears altogether.
At low~$\lambda$ (panel~a) this effect is most pronounced:
the amplified feedback increases
$\tilde{d}_s^{\mathrm{dist}}$ toward
$\tilde{d}_s^{\mathrm{raw}}$, reducing the advantage~$\rho^2$ (Eq.~\eqref{eq:cost_ratio}).
Once the converged ratio
$\tilde{\rho} \equiv
 \tilde{d}_s^{\mathrm{dist}}/\tilde{d}_s^{\mathrm{raw}}$
satisfies
$(n_{\mathrm{pairs}}/p_{\mathrm{succ}})\,\tilde{\rho}^2 > 1$
across the entire feasible~$F_0$ range,
no-distillation is everywhere cheaper and no crossover
remains. Conversely, increasing~$\lambda$ shrinks the collection window
$C^{\mathrm{round}}/\lambda$; as
$\eta_{\mathrm{link}} \to \infty$ the dashed curves collapse onto
the solid (static) ones, recovering the static limit.
The static cost ranking is therefore a conservative guide for
protocol selection: whenever it favours no-distillation, the
decoherence analysis confirms or strengthens
that conclusion.

\paragraph{Minimum link efficiency threshold\label{subsubsec:minimum_link}} For a given $\tau_{\mathrm{coh}}$, achieving
$p_L^{\mathrm{target}}$ requires a minimum link efficiency set by a
self-consistency condition. During collection of one round's Bell
pairs, the earliest pair decays to
$F_{\mathrm{stored}} \approx
F_0\,e^{-C^{\mathrm{round}}/\eta_{\mathrm{link}}}$
(Eq.~\eqref{eq:fidelity_S1}), where $C^{\mathrm{round}}$ is the
per-round raw consumption
(Eq.~\eqref{eq:round_raw}, Eq.~\eqref{eq:round_dist}). This
degradation increases the code distance needed to meet
$p_L^{\mathrm{target}}$, which increases $C^{\mathrm{round}}$,
which lengthens collection and causes further decay.

Throughout the self-consistency analyses of this work, the tilde
marks quantities determined by the iteration: $\tilde{d}_s$ is the self-consistent code distance, and $\tilde{p}_{\mathrm{Bell}}
(\tilde{d}_s)$,
$\tilde{p}_{\mathrm{local}}(\tilde{d}_s)$ (introduced in
Sec.~\ref{app:scheduling_comparison}) are the corresponding
effective Bell-pair and local error rates, distinguished from
their bare counterparts $1-F_0$ and $p_{\mathrm{phys}}$ used in
the static analysis of Sec.~\ref{subsec: fitted}.

The self-consistent code distance $\tilde{d}_s$ satisfies, neglecting idle penalty (Sec.~\ref{app:scheduling_comparison}), without
distillation,
\begin{equation}
\tilde{d}_s
= d_s^{*}\!\left(
  1 - F_0\,e^{-C_{\mathrm{raw}}^{\mathrm{round}}(\tilde{d}_s)/\eta_{\mathrm{link}}}\;
\right),
\label{eq:sc_nodist}
\end{equation}
and with distillation,
\begin{equation}
\tilde{d}_s
= d_s^{*}\!\left(
  f_{\mathcal{D}}\!\left(
    1 - F_0\,e^{-C_{\mathrm{dist}}^{\mathrm{round}}(\tilde{d}_s)/\eta_{\mathrm{link}}}
  \right)\;
\right).
\label{eq:sc_dist}
\end{equation}

Starting from the static estimate
$d_s = d_s^{*}(p_{\mathrm{raw}})$
(Eq.~\eqref{eq:dstar_def}), each iteration computes the
decay-degraded fidelity at the current $\tilde{d}_s$ and updates the
required distance. Above a critical
$\eta_{\mathrm{link}}^{*}(F_0,\, p_L^{\mathrm{target}})$, the
iteration converges to a finite $\tilde{d}_s \geq d_s$. Below this threshold, the iteration diverges, and no finite
$\tilde{d}_s$ satisfies Eq.~\eqref{eq:sc_nodist}: the target
$p_L^{\mathrm{target}}$ is unachievable regardless of
scheduling strategy.
We emphasize that this self-consistency loop is solved at
design time. It determines the code distance that must be
chosen before the lattice-surgery operation begins, given
the expected decoherence during Bell-pair collection.

Equivalently, the converged solution defines a discard fidelity
$F_{\mathrm{discard}}(\tilde{d}_s, p_L^{\mathrm{target}})$, the
lowest fidelity at which $\tilde{d}_s$ still achieves
$p_L^{\mathrm{target}}$. The self-consistent bound then takes the
form of a production constraint at this fidelity,
\begin{equation}
\eta_{\mathrm{link}}
\;\geq\;
\frac{C^{\mathrm{round}}(\tilde{d}_s)}
     {\ln(F_0/F_{\mathrm{discard}}(\tilde{d}_s, p_L^{\mathrm{target}})}
\left(1 + \frac{\Phi^{-1}(0.99)}
               {\sqrt{C^{\mathrm{round}}(\tilde{d}_s)}}\right).
\label{eq:eta_production}
\end{equation}
A narrow marginal regime exists immediately below this bound.
The bound incorporates a conservative 99th-percentile collection-time
margin. Below it, this guarantee no longer holds; operation may
nonetheless remain viable when the number of non-expired pairs within
a sliding window of duration $t_{\mathrm{discard}}$ meets the per-round
requirement, whether through a marginally sufficient mean generation
rate or transient arrival bursts. Such operation is sustained by
over-generation, at the expense of increased memory occupancy, and we
do not consider this regime further.

\subsection{\label{sec:app} Integrated feasibility criteria and platform assessment}

We briefly consider the feasibility of our methods in the context of current and next-generation neutral atom and ion trap quantum computers. A more detailed analysis requires a complete description of the intra-module architectures. 

\textit{Trapped ions.}\; 
We analyze which operating regime is accessible for state-of-the-art
trapped-ion systems. The highest demonstrated remote Bell-pair fidelity
$F_0 = 97.0\%$~\cite{Saha_2025}, achieved via time-bin photonic
encoding, falls comfortably within the no-distillation regime identified
in Sec.~\ref{sec: static}, where direct consumption of raw Bell
pairs is resource-optimal, but at a very limited generation rate of $\lambda = 0.35\,\mathrm{s}^{-1}$. The highest demonstrated entanglement generation rate
$\lambda = 250\,\mathrm{s}^{-1}$, achieved via polarization encoding at $F_0 = 94\%$ ~\cite{PhysRevLett.133.090802}, represents a substantially higher rate, yet still falls short of the on-the-fly threshold by roughly two orders of magnitude even at the least strict target of $p_L = 10^{-3}$, assuming a conservative $\tau_{\mathrm{SE}} = 1\,\mathrm{ms}$~\cite{leone2024resourceoverheadsattainablerates}, placing current trapped-ion systems in the no-expire regime. Note that the high-rate and high-fidelity results stem from different experiments employing distinct photonic encodings. Reaching the on-the-fly regime will likely require spatial multiplexing of photonic interfaces, where deploying $I > 1$ parallel optical channels per module scales $\lambda$ linearly with $I$.

Currently, entanglement stored on a memory qubit retains a fidelity
of $0.81$ after $10\,\mathrm{s}$~\cite{PhysRevLett.130.090803},
corresponding to $\tau_{\mathrm{coh}} \approx 65\,\mathrm{s}$ via
Eq.~\eqref{eq:fidelity_decay}, giving $\eta_{\mathrm{link}} =
\lambda\,\tau_{\mathrm{coh}} \approx 1.6 \times 10^4$.
While fundamental single-ion memories have demonstrated coherence
times exceeding $5500\,\mathrm{s}$~\cite{Wang_2021}, bridging this gap in network operations would theoretically project $\eta_{\mathrm{link}}$ to the order of $10^6$. 

The binding constraint is processor capacity (see Supplementary Information \ref{app:phyQ}): a single $d_s = 5$ logical qubit requires $2d_s^2 - 1 = 49$ physical qubits. The quantum charge-coupled device (QCCD) architecture \cite{Pino_2021} is a promising platform for intra-module scaling, with state-of-the-art processors recently reaching 98 physical qubits \cite{Dasu_2026}. High-rate codes can partly relax this constraint: the
$[\![80,48,4]\!]$ concatenated iceberg code, for
example, encodes 48 logical qubits on the same
processor~\cite{Dasu_2026}.  However, achieving lower logical error rates requires additional
concatenation levels, which multiplies the physical-qubit count
per logical qubit. In a distributed
setting, the low pseudo-threshold of such codes
(${\sim}4\times10^{-3}$ in circuit-level simulations,
roughly an order of magnitude below that of the
surface code) and their reliance on dense intra-block
connectivity further limit their applicability.

\textit{Neutral atoms.}
Atom--photon entanglement in neutral-atom systems has
progressed steadily over the past two
decades.
For single-node atom--photon entanglement,
the highest reported raw fidelity is $F_0 = 0.952$ using
free-space polarization-encoded collection~\cite{Zhang_2022}, while cavity-enhanced
collection has achieved a single-attempt
success probability of $0.33$ at
$F_0 = 0.866(50)$~\cite{Hartung_2024}.
Most recently, a compact parabolic-mirror node reached $F_0 = 0.93$ with a per-attempt
success probability of $0.022$~\cite{safari2026efficientcompactquantumnetwork}, and
telecom-wavelength time-bin encoding with $^{171}$Yb yielded
$F_0 = 0.90(1)$~\cite{Li_2025}.

For two-node atom--atom remote entanglement, heralded Bell pairs have been distributed over a $400 $m line-of-sight link ($700 $m fibre) at $F_0 \geq 0.892(23)$~\cite{Zhang_2022}, with heralding rates of order $10^{-2}\mathrm{s}^{-1}$, still several orders of
magnitude below the on-the-fly threshold. 
Projected cavity-enhanced schemes with
$\lambda \sim 10^5\,\mathrm{s}^{-1}$ at
$F_0 \approx 0.999$~\cite{PRXQuantum.5.020363}, combined
with the observed $\tau_{\mathrm{coh}} \sim 10\,\mathrm{s}$-scale hyperfine
coherence~\cite{Manetsch:2024lwl}, would yield
$\eta_{\mathrm{link}} \gtrsim 10^6$, placing neutral-atom
links well within the on-the-fly correction regime. At the same time, recent experiments have demonstrated that
neutral-atom network nodes support strong spatial  multiplexing, enabling single nodes to achieve
substantially enhanced entanglement generation rates that
scale with the number of emitters~\cite{Li_2025,Hartung_2024}.

The potential advantage over trapped ions is module capacity: arrays of
several hundred physical qubits~\cite{Bluvstein_2023,
Manetsch:2024lwl} relax the processor-size constraint that
limits current ion-trap modules.

%\sitong{remove?}
%\textit{Color centers.}\; We estimate $\eta_{\mathrm{link}}$ for state-of-the-art color-center systems by combining the best capabilities across different point defects. For local processing, NV-center nodes have demonstrated registers of $N_{\mathrm{phy}} \approx 10$ physical qubits (1 electron and 9 nuclear-spin memories)~\cite{Bradley_2019}; however, this still falls short of the number of physical qubits required for a $d_s = 5$ surface-code patch.

%For network connectivity, SiV-center nodes currently provide the benchmark. The best demonstrated remote electron--electron Bell-pair rate is $\lambda \approx 1\,\mathrm{s}^{-1}$ at $F_0 = 0.86(3)$. When stored into the local $^{29}$Si nuclear memory ($\tau_{\mathrm{coh}} \approx 2\,\mathrm{s}$, Table~S4 of~\cite{Knaut_2024}), this gives a link efficiency of $\eta_{\mathrm{link}} = \lambda\,\tau_{\mathrm{coh}} \approx 2$, six orders of magnitude below the on-the-fly threshold ($\eta_{\mathrm{link}} \sim 10^6$) required for scalable distributed error correction.

\section{\label{sec:discussion} Discussion}

We have analyzed the resource trade-off between
distillation-free and distillation-assisted remote lattice
surgery for the rotated surface code under both static and
time-dependent noise models. Three main findings emerge.

First, we derive an explicit condition
(Eq.~\eqref{eq:nodist_optimal}) for choosing between
direct and distillation-assisted consumption of Bell pairs:
distillation-free operation is resource-optimal when the
quadratic distance saving from purification no longer
compensates for the distillation overhead.
Under a static model with no memory decoherence, this
crossover occurs at $F_0 \approx 95\,\%$, is stable across
$p_L \in [10^{-3},\, 10^{-12}]$, and yields overhead
reductions up to $68\,\%$
($F_0 \approx 98.64\,\%$, $p_L = 10^{-3}$).

Second, incorporating finite generation rates and memory
decoherence, we identify three operating regimes:
on-the-fly, where generation keeps pace with each
syndrome round; no-expire, where pairs arrive slower and must wait in
memory, but their fidelity remains usable by the time
they are consumed; and infeasible, where decoherence outpaces any achievable code
distance. In the no-expire regime, Bell-pair decay and
data-qubit idle errors feed back into the required code
distance via a self-consistency condition
(Eq.~\eqref{eq:self_consistent_ds}), inflating costs
beyond the static estimate. Nonetheless, where
distillation-free operation is already favored under the
static model, this conclusion holds under time-dependent decoherence.

Third, the best demonstrated trapped-ion fidelity
($F_0 = 97\,\%$~\cite{Saha_2025}) already falls within
the distillation-free regime, while the highest
demonstrated rate
($\lambda = 250\,\mathrm{s}^{-1}$ at $F_0 = 94\,\%$~\cite{PhysRevLett.133.090802})
places current systems in the no-expire window.
Projected neutral-atom links
($\lambda \sim 10^5\,\mathrm{s}^{-1}$,
$F_0 \approx 0.999$~\cite{PRXQuantum.5.020363})
would reach the on-the-fly regime, though demonstrated
rates remain much lower~\cite{Zhang_2022}. These two platforms are representative of the parameter
space analyzed here and provide practical co-design targets
for photonic interconnects, memory lifetimes, and
fault-tolerant logical layouts.
An alternative approach is transversal gate
teleportation, discussed in
Supplementary Information~\ref{app:transversal}.

Several natural extensions of this analysis remain for future work. The scheduling analysis in this work focuses on single lattice surgery operations and does not consider idle periods between successive remote operations to pre-generate Bell pairs; accounting for algorithm-level scheduling and inter-module connectivity could further reduce overhead. The analysis presented here is platform-independent, but it can be combined with platform-specific parameters to provide a foundation for systematic architectural design; a
detailed resource analysis along these lines will be
presented in forthcoming work. Finally, this study assumes
a uniform code distance throughout the architecture;
allowing different distances for data, routing, and
communication patches may further reduce the physical
qubit overhead.

\section{Methods}
\subsection{\label{sec:FAULT-TOLERANT} Fault-tolerant requirements for remote lattice surgery}

We consider a distributed architecture in which modules are
interconnected by photonic links, and each module hosts logical
qubits encoded in distance-$d$ rotated surface codes.

\subsubsection{Remote lattice surgery protocol}
\label{subsec:patch_layout}

Logical gates in the rotated surface-code lattice-surgery
framework are realized through Pauli product measurements
(PPMs): adjacent patches (distance $d$, $d^2$ physical qubits
each) can be merged directly along their shared boundary (the \emph{seam}), or connected through an ancilla patch for multi-qubit and
long-range operations. In either case, a
merge--measure--split sequence with $d$ syndrome rounds of
stabilizer measurement is required to extract the target Pauli product operator at full code distance. The measurement projects the
participating qubits onto the $+1$ or $-1$ eigenspace of the
target multi-qubit Pauli product operator, with measurement
outcomes determining Pauli corrections that are tracked
classically. In the Pauli-based computation framework, all
Clifford+$T$ circuits reduce to a sequence of such PPMs, where
each non-Clifford $T$ gate consumes one magic state and one
PPM~\cite{Litinski2019gameofsurfacecodes}. We refer the reader
to Refs.~\cite{horsman-patch-surface-code,
Litinski2018latticesurgery, Litinski2019gameofsurfacecodes, 10313779} for a detailed description.

%Logical two-qubit gates are realized via the rotated surface-code lattice-surgery framework~\cite{horsman-patch-surface-code, Litinski2019gameofsurfacecodes}: joint logical Pauli operators between code patches (distance $d$, $d^2$ physical qubits each) are measured through a merge--measure--split sequence~\cite{Fowler_2012, PhysRevA.90.062320}. The merge step introduces interface stabilizers at the patch boundary (\emph{seam}), measured over $d_s$ consecutive syndrome rounds to preserve the code distance.

For intra-module operations, seam stabilizers are measured with local gates.
For inter-module operations, seam stabilizers span physically separated
hardware, and remote parity checks require Bell pairs shared between
modules~\cite{sunami2025entanglementboostinglowvolumelogical,
jacinto2025networkrequirementsdistributedquantum}.

\subsubsection{Bell-pair generation: mechanism and rates}
\label{subsec:entanglement_generation}

Remote entanglement is established via \emph{heralded photonic
links}~\cite{Moehring:2007ywt,Monroe_2014,PhysRevLett.124.110501,PRXQuantum.2.017002}: a coincident detection event at a
Bell-state analyzer heralds successful projection into an entangled Bell
state~\cite{Humphreys2017DeterministicDO}, with overall success
probability $p_{\mathrm{herald}} \ll 1$.

The post-heralded fidelity $F_{\mathrm{Bell}}$ is governed by distinct
physical mechanisms: Hong-Ou-Mandel visibility, channel losses, detector
dark counts~\cite{Dhara}, and qubit decoherence during the heralding
window~\cite{PhysRevLett.124.110501}. Distributed Bell pairs
are typically an order of magnitude noisier than local
operations~\cite{Saha_2025,PhysRevLett.117.060504}.
We treat $F_0 = 1 - p_{\mathrm{raw}}$ as a configurable input informed by experimental
benchmarks rather than deriving it from physical-layer parameters.

Scalability is jointly limited by two hardware
characteristics: (i)~the \emph{generation rate} $\lambda$  and (ii)~the \emph{fidelity} $F_0$, which must exceed the fault-tolerance threshold. Each module carries $I$
optical interfaces~\cite{PRXQuantum.2.020331, Takahashi_2020, PRXQuantum.2.017002}
which together determine its Bell-pair generation rate $\lambda$.

\subsubsection{Seam error threshold}
\label{subsec:Bell-pair threshold}

Recent studies have characterized the fault-tolerance properties of seam operations in distributed surface codes. Numerical simulations reveal that syndrome extraction at the seam tolerates error rates approximately one order of magnitude higher than bulk operations~\cite{noisylink2024}. The
underlying mechanism is that errors from noisy Bell pairs
propagate only along a single spatial dimension of the surface code during lattice surgery, limiting the number of paths through which errors can form logical failures~\cite{noisylink2024}.

For the unrotated surface code under a phenomenological noise model,
the thresholds are approximately $p_{\mathrm{th}}^{\mathrm{bulk}} \approx 1\%$
and $p_{\mathrm{th}}^{\mathrm{seam}} \approx 10\%$~\cite{noisylink2024}.
For the rotated surface code (Figure~\ref{fig:remote_surgery}(b)),
circuit-level simulations confirm similar threshold
behavior~\cite{shalby2025optimizednoiseresilientsurfacecode,jacinto2025networkrequirementsdistributedquantum,haug2025latticesurgerybellmeasurements}.
Under an amplification factor
$\Gamma = p_{\mathrm{th}}^{\mathrm{seam}}/p_{\mathrm{th}}^{\mathrm{bulk}} = 10$,
fault tolerance for gate-teleportation interfaces persists with
$p_{\mathrm{th}}^{\mathrm{bulk}} \approx 0.54\%$ and
$p_{\mathrm{th}}^{\mathrm{seam}} \approx 5.4\%$
~\cite{shalby2025optimizednoiseresilientsurfacecode}. Other teleportation protocols report Bell-pair error thresholds of
$p_{\mathrm{th}}^{\mathrm{Bell}} \approx 17\%$~\cite{haug2025latticesurgerybellmeasurements}
and $p_{\mathrm{th}}^{\mathrm{Bell}} \approx 30\%$~\cite{jacinto2025networkrequirementsdistributedquantum},
with corresponding local gate error rates of $0.1\%$ (operating point)
and $0.7\%$ (threshold), respectively.

\begin{table*}[ht]
\caption{\textbf{Noise model parameters for local and non-local
operations in distillation protocols and remote lattice surgery
(see Fig.~\ref{fig:distance_four_panels} and  Fig.~\ref{fig:cost_four_panels}).}
Single qubit idle errors apply only when data qubits stall between syndrome
rounds (Strategy~1, no-expire regime;
Sec.~\ref{subsec:decay_choice},
\ref{app:scheduling_comparison}).
}
\label{tab:noise_parameters}
\begin{ruledtabular}
\begin{tabular}{llc}
\textbf{Operation} & \textbf{Noise model}
  & \textbf{Error rate} \\
\hline
\multicolumn{3}{l}{\textbf{Local operations}} \\
Single-qubit gates
  & Depolarizing: uniform Pauli error over $\{X,Y,Z\}$,
    each $p_{\mathrm{local}}/3$
  & $p_{\mathrm{local}} = 0.1\%$ \\
Two-qubit gates
  & Two-qubit depolarizing: uniform over 15 non-identity
    Paulis, each $p_{\mathrm{local}}/15$
  & $p_{\mathrm{local}} = 0.1\%$ \\
Measurement / Reset
  & Bit-flip (measurement-basis–specific)
  & $p_{\mathrm{local}} = 0.1\%$ \\
Single-qubit idle errors
  & Depolarizing: uniform Pauli error over $\{X,Y,Z\}$,
    each $p_{\mathrm{idle}}/3$
  & $p_{\mathrm{idle}}$ (Eq.~\eqref{eq:idle_error}) \\
\hline
\multicolumn{3}{l}{\textbf{Non-local operations}} \\
Bell-pair preparation
  & Two-qubit depolarizing on distributed Bell state
  &$p_{\mathrm{Bell}} = 1 - F$ \\
\end{tabular}
\end{ruledtabular}
\end{table*}

\subsection{Resource overhead under static fidelity}
\subsubsection{Fitted model for code distance requirements}
\label{subsec: fitted}
To quantify the relationship between Bell-pair fidelity and surface-code distance, we adopt the logical error model and fitted values from Sunami \textit{et al.}~\cite{sunami2025entanglementboostinglowvolumelogical}. For a distance-$d_s$ rotated surface code undergoing remote lattice surgery with Bell-pair error rate $p_{\mathrm{Bell}}$ and local gate error rate $p_{\mathrm{local}}$, the logical error rate is given by

\begin{equation}
\label{eq:bell-logical-error}
\begin{aligned}
p_L
&= \kappa (d_s + 1)^{\eta}
\Big[
A^{\frac{d_s+1}{2}}
+ B^{\frac{d_s+1}{2}} \\
&\quad
+ \sum_{\gamma_s=1}^{d_s}
(A M^2)^{\gamma_s/2}
B^{\frac{d_s+1-\gamma_s}{2}}
\Big],
\end{aligned}
\end{equation}
where $A = p_{\mathrm{Bell}}/p_{\mathrm{th}}^{\mathrm{Bell}}$, $B = p_{\mathrm{local}}/p_{\mathrm{th}}^{\mathrm{local}}$, and $M = 1 + \alpha_c p_{\mathrm{local}} p_{\mathrm{th}}^{\mathrm{Bell}}/(1-\sqrt{B})$. The model parameters, obtained from circuit-level simulations, are $\kappa = 5.44 \times 10^{-2}$, $\eta = 5.34 \times 10^{-1}$, $\alpha_c = 3.15 \times 10^{2}$, $p_{\mathrm{th}}^{\mathrm{Bell}} = 15.3\%$, and $p_{\mathrm{th}}^{\mathrm{local}} = 1.02\%$. Note that in the full model, the cross terms in Eq.~\eqref{eq:bell-logical-error} reduce the effective Bell-pair error threshold to $p_{\mathrm{Bell}}^{\mathrm{eff}} = p_{\mathrm{th}}^{\mathrm{Bell}}/M^2 \approx 13.36\%$ at $p_{\mathrm{local}}=0.1\%$ ($F_0 \approx 86.64\%$), below the nominal fitted value of $15.3\%$.

For a target logical error rate $p_L^{\mathrm{target}}$ and a given local
error rate $p_{\mathrm{local}}$ (fixed to $0.1\%$ throughout), we define the
minimum required surface-code distance as
\begin{equation}
d_s^{*}(p_{\mathrm{Bell}})
=
\min_{\substack{d_s \in 2\mathbb{N}+1 \\ d_s \geq 3}}
\Bigl\{
d_s :
p_L(d_s, p_{\mathrm{Bell}}, p_{\mathrm{local}})
\le
p_L^{\mathrm{target}}
\Bigr\},
\label{eq:dstar_def}
\end{equation}
where $p_L(\cdot)$ denotes the logical error-rate model. In practice, 
$d_s^{*}$ is obtained via binary search over odd code distances.

As $p_{\mathrm{Bell}}$ approaches $p_{\mathrm{th}}^{\mathrm{Bell}}$
($A\to 1$), the required $d_s$ grows rapidly. With $p_{\mathrm{local}} = 0.1\%$ and
$p_L^{\mathrm{target}}$ fixed, this scaling is determined by the pure
Bell-pair term $A^{(d_s+1)/2}$ in
Eq.~\eqref{eq:bell-logical-error}, where
$A = p_{\mathrm{Bell}}/p_{\mathrm{th}}^{\mathrm{Bell}}$.
The largest cross term ($\gamma_s = d_s$),
$(AM^2)^{d_s/2}B^{1/2}$, carries an extra factor $M^2$ in its
base relative to the pure Bell-pair term
$A^{(d_s+1)/2}\propto A^{d_s/2}$. When $p_{\mathrm{local}}$ is
small, $M$ is close to $1$, so the cross term only mildly
modifies the leading scaling set by $A^{(d_s+1)/2}$.

Inverting the leading exponential dependence gives
\begin{equation}
\label{eq:dstar_scaling}
d_s
\propto
\frac{1}
     {\log(p_{\mathrm{th}}^{\mathrm{Bell}}/p_{\mathrm{Bell}})},
\end{equation}
which diverges as
$p_{\mathrm{Bell}} \to p_{\mathrm{th}}^{\mathrm{Bell}}$.
Since the per-round Bell-pair consumption scales as
$n^{\mathrm{round}} \propto d_s$, the total resource overhead
grows correspondingly as the threshold is approached.

Although the numerical coefficients in
Eq.~\eqref{eq:bell-logical-error} are specific to the fitted
model of
Ref.~\cite{sunami2025entanglementboostinglowvolumelogical},
the exponential suppression and threshold asymmetry are generic
features confirmed by independent
simulations~\cite{noisylink2024,
jacinto2025networkrequirementsdistributedquantum}. In the
standard surface-code threshold
argument~\cite{Fowler_2012}, the logical error rate per round
scales as
$p_L \sim (p/p_{\mathrm{th}})^{\lfloor(d+1)/2\rfloor}$. The physical distinction between seam and bulk noise summarized in Sec.~\ref{subsec:Bell-pair threshold} is what
motivates the large separation
$p_{\mathrm{th}}^{\mathrm{Bell}} \gg p_{\mathrm{th}}^{\mathrm{local}}$
adopted in Eq.~\eqref{eq:bell-logical-error}.

\subsubsection{Distillation protocols and distance scaling \label{subsec:distillation}}

Entanglement distillation improves Bell-pair fidelity by
consuming multiple noisy pairs to produce a single purified
pair, enabling either the use of otherwise unsuitable raw links
or the reduction of code distance requirements for a given raw
input fidelity $F_0$. A distillation protocol $\mathcal{D}$
consumes $n_{\mathrm{pairs}}$ raw Bell pairs with error rate
$p_{\mathrm{raw}}=1-F_0$ and, with success probability
$p_{\mathrm{succ}}$, outputs one purified pair with reduced
error rate. It is characterized by two mappings:
\begin{equation}
p_{\mathrm{out}} = f_{\mathcal{D}}(p_{\mathrm{raw}}),
\qquad
p_{\mathrm{succ}} = g_{\mathcal{D}}(F_0;\, p_{\mathrm{local}}),
\end{equation}
where $p_{\mathrm{out}}$ is the output error rate, $p_{\mathrm{succ}}$
is the success probability, $F_0 = 1 - p_{\mathrm{raw}}$ is the raw
fidelity (post-heralded Bell-pair fidelity), and $p_{\mathrm{local}}$ captures local gate and measurement
errors during the distillation circuit. The input consumption $n_{\mathrm{pairs}}$ is protocol-dependent.

The effective Bell-pair error rate entering seam syndrome extraction is
\begin{equation}
p_{\mathrm{Bell}}^{\mathrm{eff}} =
\begin{cases}
p_{\mathrm{raw}}, & \text{(no distillation)}, \\
f_{\mathcal{D}}(p_{\mathrm{raw}}), & \text{(with protocol } \mathcal{D}\text{)},
\end{cases}
\label{eq:p_out}
\end{equation}
denoted $p_{\mathrm{eff}}$ for brevity. Since the logical error model
Eq.~\eqref{eq:bell-logical-error} is monotonically non-decreasing in
$p_{\mathrm{Bell}}$, the required code distance $d_s^{*}(p_{\mathrm{Bell}})$
inherits this monotonicity. Any effective distillation protocol
therefore satisfies $p_{\mathrm{eff}} < p_{\mathrm{raw}}$ and enables
a distance reduction
\begin{equation}
\Delta d_s = d_s^{*}(p_{\mathrm{raw}}) - d_s^{*}(p_{\mathrm{eff}}) \geq 0,
\label{eq:delta_d}
\end{equation}
decreasing the per-round Bell-pair consumption from
$a d_s^{*}(p_{\mathrm{raw}}) - c$ to $a d_s^{*}(p_{\mathrm{eff}}) - c$.
We quantify this reduction through the distance ratio
\begin{equation}
\rho(p_{\mathrm{raw}})
= \frac{d_s^{*}(p_{\mathrm{eff}})}{d_s^{*}(p_{\mathrm{raw}})},
\label{eq:distance_ratio}
\end{equation}
where $\rho \in (0,1)$, with smaller values indicating greater distance
savings. This per-round reduction must however be weighed against the
multiplicative overhead $n_{\mathrm{pairs}}/p_{\mathrm{succ}}$ required
to produce each purified pair. The quantitative distance reductions
achieved by representative protocols are shown in
Fig.~\ref{fig:distance_four_panels}. Since $d_s$ grows rapidly as $F_0$ approaches the seam
threshold (Sec.~\ref{subsec: fitted}) and becomes impractically
large at lower $F_0$, we restrict our analysis to
$F_0 \ge 90\%$, which covers Bell-pair fidelities reported in
photon-mediated remote-entanglement experiments to date
(Sec.~\ref{sec:app}).

\subsubsection{Expected Bell-pair consumption under probabilistic distillation}

Entanglement distillation reduces the required code distance from
$d_s^{*}(p_{\mathrm{raw}})$ to $d_s^{*}(p_{\mathrm{eff}})$, lowering
the per-round purified pair requirement to
$n^{\mathrm{round}}(p_{\mathrm{eff}}) = a d_s^{*}(p_{\mathrm{eff}}) - c$.
However, producing each purified pair requires $n_{\mathrm{pairs}}$ raw
pairs per attempt and succeeds only with probability $p_{\mathrm{succ}}
= g_{\mathcal{D}}(F_0; p_{\mathrm{local}})$, so the total raw
consumption carries a multiplicative overhead.

Distillation attempts can be executed either serially or in parallel;
detailed analysis of both execution models is provided in
Supplementary Information~\ref{app:distillation_execution}. For a fixed module pair, with $R$ the total syndrome rounds
across consecutive lattice-surgery operations and distilled
pairs reusable across rounds, the per-round raw consumption
converges asymptotically to
\begin{equation}
\lim_{R \to \infty} \frac{\mathbb{E}[N_{\mathrm{raw}}^{\mathrm{total}}]}{R}
= \frac{n_{\mathrm{pairs}}}{p_{\mathrm{succ}}} \cdot n^{\mathrm{round}},
\label{eq:distillation_rate}
\end{equation}
independent of execution mode. Terminal overhead from unused buffered pairs is negligible for
$R$ large compared to the memory size. The long-term Bell-pair cost per round is therefore
\begin{equation}
C_{\mathrm{dist}}^{\mathrm{round}}
= \frac{n_{\mathrm{pairs}}}{p_{\mathrm{succ}}}
\times n^{\mathrm{round}}(p_{\mathrm{eff}}),
\label{eq:cost_longterm}
\end{equation}
where $n^{\mathrm{round}}(p_{\mathrm{eff}}) = a d_s^{*}(p_{\mathrm{eff}})
- c$ and $p_{\mathrm{eff}} = f_{\mathcal{D}}(p_{\mathrm{raw}})$.

\subsection{Bell-pair fidelity decay model}
\subsubsection{Stochastic Bell-pair generation}

When Bell-pair generation proceeds via repeated heralded attempts with success
probability $p_{\mathrm{herald}} \ll 1$, the sequence of successful events is
well-approximated by a Poisson process~\cite{PhysRevLett.133.090802,PhysRevLett.124.110501,Hucul2014ModularEO}
with rate $\lambda = I \cdot r_{\mathrm{attempt}} \cdot p_{\mathrm{herald}}$ (Hz),
where $I$ optical interfaces each attempt at rate $r_{\mathrm{attempt}}$. The 
inter-arrival time between consecutive successful generations is exponentially 
distributed with mean $1/\lambda$.

\textit{Collection time statistics.}
Collecting $n^{\mathrm{round}}$ pairs requires waiting through $n^{\mathrm{round}}$
independent inter-arrival times. The total collection time $T_{\mathrm{total}}$ 
follows an Erlang distribution with mean and variance
\begin{equation}
\mathbb{E}[T_{\mathrm{total}}] = \frac{n^{\mathrm{round}}}{\lambda}, 
\quad
\mathrm{Var}[T_{\mathrm{total}}] = \frac{n^{\mathrm{round}}}{\lambda^2}.
\end{equation}

For sufficiently large $n^{\mathrm{round}}$, $T_{\mathrm{total}}$ is approximately 
Gaussian by the central limit theorem. The 99th-percentile collection time is then
\begin{equation}
t_{0.99} = \frac{n^{\mathrm{round}}}{\lambda}
\left(1 + \frac{\Phi^{-1}(0.99)}{\sqrt{n^{\mathrm{round}}}}\right),
\label{eq:t99}
\end{equation}
where $\Phi^{-1}(0.99) \approx 2.33$ is the 99th percentile of the standard
normal distribution. For typical code distances $d_s \in \{5,7,9,11\}$ ($n^{\mathrm{round}} \sim 9$--$21$), 
the Gaussian approximation error in the 99th-percentile
collection time is ${\sim}4$-$8\%$, and sufficient for system-level estimates.

\subsubsection{Storage decay}
\label{subsub:decay}
Successfully heralded Bell pairs decohere during storage as
\begin{equation}
F(t) = F_0 \, e^{-t/\tau_{\mathrm{coh}}},
\label{eq:fidelity_decay}
\end{equation}
where $F_0$ is the initial fidelity and $\tau_{\mathrm{coh}}$ is the 
memory coherence time. This is a high-fidelity approximation to the 
exact depolarizing model $F(t) = \tfrac{1}{4} + \bigl(F_0 - \tfrac{1}{4}\bigr)\,e^{-t/\tau_{\mathrm{coh}}}$,
valid when $F(t) \gg 1/4$.

Since pairs in a batch arrive at different times, they experience different
storage durations before consumption. We impose a \textit{storage cutoff
time}~\cite{Leveraging,10901810, grimbergen2026probabilisticcutoffshomogeneousquantum}: any pair is discarded if its storage time would increase its error rate above
$p_{\mathrm{discard}} = 13.3\%$, chosen slightly below the effective
threshold observed in our finite-size simulations,
$p_{\mathrm{th}} \approx 13.36\%$.
Evaluating batch quality conservatively via the worst-case (first-generated) pair,
the condition $F_0 e^{-t/\tau_{\mathrm{coh}}} \ge F_{\mathrm{discard}}$
yields the maximum viable storage time
\begin{equation}
t_{\mathrm{discard}} = \tau_{\mathrm{coh}}
\ln\!\left(\frac{F_0}{F_{\mathrm{discard}}}\right).
\label{eq:t_discard}
\end{equation}

\subsection{Detailed comparison of scheduling strategies}
\label{app:scheduling_comparison}

A critical bottleneck in remote lattice surgery arises when the entanglement generation rate $\lambda$ cannot sustain continuous lattice
surgery operations, i.e., when the mean rate $\lambda < n^{\mathrm{round}}/T^{\mathrm{round}}$ (Sec.~\ref{subsec:decay_choice}). The system must then absorb unavoidable idle time, introducing a
fundamental scheduling dilemma~\cite{Bonilla_Ataides_2025}.
Accumulating all $N_{\mathrm{QEC}}$ pairs before lattice surgery exposes early
Bell pairs to severe memory decay, whereas executing surgery round-by-round
with intermediate pauses exposes the coupled data qubits to
idle errors that accumulate between syndrome measurement rounds. We analyze two concrete strategies that
represent these extremes.

\textbf{Strategy 1: Round-by-round Scheduling}
Generate $n^{\mathrm{round}} = a d_s - c$ Bell pairs per syndrome round,
then immediately execute one round of syndrome extraction with
teleported CNOTs along the seam. Because
$\lambda < n^{\mathrm{round}}/T^{\mathrm{round}}$, both patches idle
with the noisy seam coupled while awaiting the next batch, and the
cycle repeats across $d_s - 1$ rounds. Neglecting stochastic
fluctuations in the Poisson arrival times (see Eq.~\eqref{eq:t99}),
the earliest pair in each batch waits approximately
$n^{\mathrm{round}}/\lambda$ and decays to
\begin{equation}
\label{eq:fidelity_S1_app}
F_{\mathrm{stored}}^{\mathrm{S1}}
    \approx F_0 \exp\!\left(-\frac{n^{\mathrm{round}}}
    {\eta_{\mathrm{link}}}\right),
\end{equation}
giving $p_{\mathrm{Bell}}^{\mathrm{S1}}
= 1 - F_{\mathrm{stored}}^{\mathrm{S1}}$.
Between rounds, data qubits wait
$\Delta t_{\mathrm{idle}} \approx n^{\mathrm{round}}/\lambda$ with
the seam coupled. Modeling this wait as symmetric depolarizing noise with characteristic
time $\tau_{\mathrm{dep}} = \mu\,\tau_{\mathrm{coh}}$, the per-round idle
error applied to every data qubit is
\begin{equation}
\label{eq:idle_error}
p_{\mathrm{idle}}
= 1 - \exp\!\left(
      -\frac{n^{\mathrm{round}}}
      {\mu\,\eta_{\mathrm{link}}}\right),
\qquad
\mu \equiv \tau_{\mathrm{dep}}/\tau_{\mathrm{coh}},
\end{equation}
where $\mu \equiv \tau_{\mathrm{dep}}/\tau_{\mathrm{coh}} \geq 2$,
since a Bell pair undergoes independent depolarization on each
half, doubling its decay rate relative to a single data qubit.
In practice $\mu$ varies when communication and data qubits
differ in species or coherence time. When $\mu\gg2$ the idle
penalty is negligible and Strategy~1 dominates; as $\mu\to2$ the
idle cost grows and Strategy~2 becomes competitive. The effective local error rate entering the decoder is
$\tilde{p}_{\mathrm{local}}
  \simeq p_{\mathrm{phys}} + p_{\mathrm{idle}}$
(neglecting the $O(p^2)$ cross-term),
where $p_{\mathrm{phys}}$ is the local error rate from all
non-idle sources (Table~\ref{tab:noise_parameters}).

The effective Bell-pair error  $\tilde{p}_{\mathrm{Bell}}$ and idle-induced
local error $\tilde{p}_{\mathrm{local}}$ both depend on $\tilde{d}_s$ through
$n^{\mathrm{round}} = a\tilde{d}_s - c$
(Eq.~\eqref{eq:round_pairs}), while $\tilde{d}_s$ is itself
determined by these error rates via Eq.~\eqref{eq:dstar_def}.
This defines a self-consistency condition analogous to
Eqs.~\eqref{eq:sc_nodist}--\eqref{eq:sc_dist},
which we solve via the iterative procedure in
Algorithm~\ref{alg:self_consistent}:
\begin{equation}
\tilde{d}_s = d_s^{*}\!\Bigl(
  \tilde{p}_{\mathrm{Bell}}(\tilde{d}_s),\;
  \tilde{p}_{\mathrm{local}}\bigl(\tilde{d}_s\bigr)
\Bigr).
\label{eq:self_consistent_ds}
\end{equation}

\textbf{Strategy 2: Pre-buffered}
Accumulate all $N_{\mathrm{QEC}} = d_s\, n^{\mathrm{round}}$ pairs
while both patches run independent local QEC cycles with no seam
coupling, then execute all $d_s$ rounds back-to-back. Because the
patches are decoupled throughout accumulation, we assume
$\tilde{p}_{\mathrm{local}}^{\mathrm{S2}} = p_{\mathrm{phys}}$ with
no idle penalty. However, the earliest pair waits $d_s$ times
longer than in Strategy~1 and decays to
\begin{equation}
\label{eq:fidelity_S2}
F_{\mathrm{stored}}^{\mathrm{S2}}
    \approx F_0 \exp\!\left(-\frac{d_s\, n^{\mathrm{round}}}
    {\eta_{\mathrm{link}}}\right),
\end{equation}
yielding $p_{\mathrm{Bell}}^{\mathrm{S2}} \gg p_{\mathrm{Bell}}^{\mathrm{S1}}$
(Eqs.~\eqref{eq:fidelity_S1},~\eqref{eq:fidelity_S2}).

Analogously, Strategy~2's effective Bell-pair error
$\tilde{p}_{\mathrm{Bell}}^{\mathrm{S2}}
$
(Eq.~\eqref{eq:fidelity_S2}) couples to $\tilde{d}_s$ through
the full-cycle consumption
$N_{\mathrm{QEC}}=\tilde{d}_s\,n^{\mathrm{round}}$ rather than
the per-round $n^{\mathrm{round}}$ of Strategy~1, while
$\tilde{p}_{\mathrm{local}}^{\mathrm{S2}}=p_{\mathrm{phys}}$ is
$\tilde{d}_s$-independent, yields 
\begin{equation}
\tilde{d}_s = d_s^{*}\!\Bigl(
  \tilde{p}_{\mathrm{Bell}}^{\mathrm{S2}}(\tilde{d}_s),\;
  p_{\mathrm{phys}}
\Bigr),
\end{equation}

\begin{figure*}[ht]
\centering
{\includegraphics[width=0.49\textwidth]{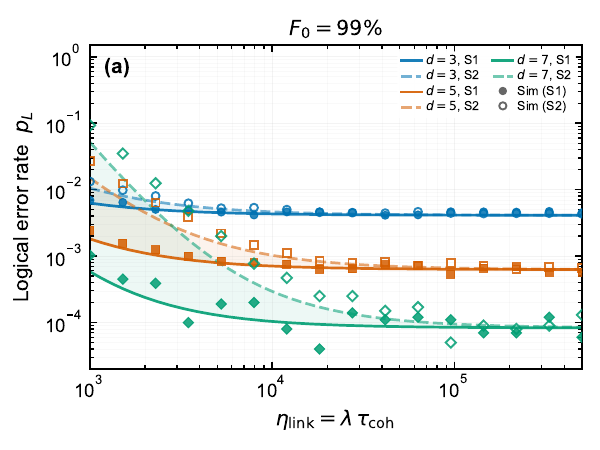}
    \label{fig:crossover_a}}
\hfill
{\includegraphics[width=0.49\textwidth]{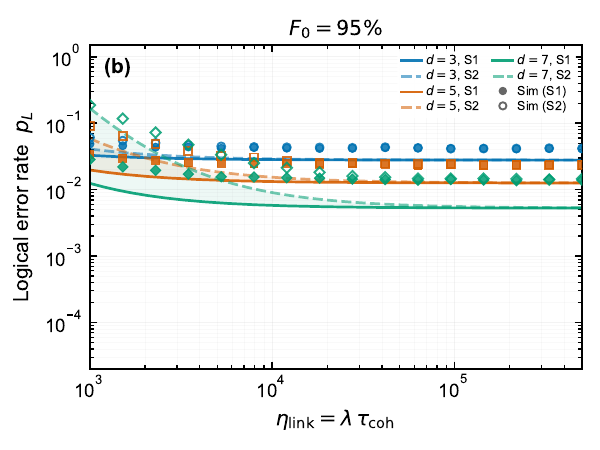}
    \label{fig:crossover_b}}
\caption{\textbf{Logical error rate versus link efficiency
  $\boldsymbol{\eta}_{\boldsymbol{\mathrm{link}}}$ for Strategies~1 and~2 at
  $\boldsymbol{d_s \in \{3,5,7\}}$.}
  Solid curves and filled markers: Strategy~1 (round-by-round);
  dashed curves and open markers: Strategy~2 (pre-buffered).
  Curves show an analytical model adapted from the fitted formula of
  Ref.~\cite{sunami2025entanglementboostinglowvolumelogical},
  with the noise parameters adapted to the teleported-CNOT error model;
  markers show Stim circuit-level simulation decoded with MWPM.
  Each simulation point is the basis-averaged logical error rate
  $p_L$,
  combining $Z_L\!\otimes\!Z_L$ and $X_L\!\otimes\!X_L$
  merge circuits ($5\times10^4$ shots per basis).
  Shaded regions highlight the S1--S2 performance gap.
  (a)~$F_0=0.99$.
  (b)~$F_0=0.95$.}
\label{fig:crossover}
\end{figure*}

\begin{algorithm}[ht]
\begin{algorithmic}[1]
\Require $\lambda$, $\tau_{\mathrm{coh}}$, $\mu$,
         $F_0$, $p_{\mathrm{phys}}$,
         $p_L^{\mathrm{target}}$,
         $d_{\max}$,
         strategy $\in \{1, 2\}$
\Ensure  $\tilde{d}_s$ or \textsc{Infeasible}

\State $\eta_{\mathrm{link}} \gets \lambda\,\tau_{\mathrm{coh}}$
\State $p_{\mathrm{Bell}} \gets 1 - F_0$
       \hfill\Comment{initial bare value}
\State $\tilde{d}_s \gets d_s^*\!\left(
       p_{\mathrm{Bell}},\; p_{\mathrm{phys}},\;
       p_L^{\mathrm{target}}\right)$
       \hfill\Comment{Eq.~\eqref{eq:dstar_def}}

\Loop
    \State $n \gets a\,\tilde{d}_s - c$
           \hfill\Comment{Eq.~\eqref{eq:round_pairs}}
    \If{strategy $= 1$}
        \State $F_{\mathrm{stored}} \gets F_0\,e^{-n/\eta_{\mathrm{link}}}$
               \hfill\Comment{Eq.~\eqref{eq:fidelity_S1_app}}
        \State $p_{\mathrm{idle}} \gets 1 - e^{-n/(\mu\,\eta_{\mathrm{link}})}$
               \hfill\Comment{Eq.~\eqref{eq:idle_error}}
    \Else
        \State $F_{\mathrm{stored}} \gets F_0\,e^{-\tilde{d}_s\,n/\eta_{\mathrm{link}}}$
               \hfill\Comment{Eq.~\eqref{eq:fidelity_S2}}
        \State $p_{\mathrm{idle}} \gets 0$
    \EndIf
    \State $\tilde{p}_{\mathrm{Bell}} \gets 1 - F_{\mathrm{stored}}$
    \State $\tilde{d}_s' \gets d_s^*\!\left(
           \tilde{p}_{\mathrm{Bell}},\;
           p_{\mathrm{phys}} + p_{\mathrm{idle}},\;
           p_L^{\mathrm{target}}\right)$
           \hfill
    \If{$\tilde{d}_s' > d_{\max}$}
        \;\Return \textsc{Infeasible}
    \EndIf
    \If{$\tilde{d}_s' = \tilde{d}_s$}
        \;\Return $\tilde{d}_s$
    \EndIf
    \State $\tilde{d}_s \gets \tilde{d}_s'$
\EndLoop
\end{algorithmic}

\caption{\textbf{Self-consistent distance solver.}
Strategy~1 (round-by-round) and Strategy~2 (pre-buffered).}
\label{alg:self_consistent}
\end{algorithm}

\paragraph{Feasibility range}
Strategy~2 requires all $N_{\mathrm{QEC}}$ pairs to survive in memory
until surgery begins, so the no-expire condition demands
\begin{equation}
\label{eq:feasibility}
\eta_{\mathrm{link}} \;\geq\; \eta_{\min}^{\mathrm{S2}}
    = \frac{N_{\mathrm{QEC}}}{\ln(F_0/F_{\mathrm{th}})},
\end{equation}
a bound $d_s$ times tighter than Strategy~1's requirement
$\eta_{\min}^{\mathrm{S1}} = n^{\mathrm{round}}/\ln(F_0/F_{\mathrm{th}})$.
Strategy~1 therefore remains viable at link efficiencies a factor of
$d_s$ below what Strategy~2 requires.

\paragraph{Performance comparison}
Where both strategies are feasible, they trade off Bell-pair
fidelity against data-qubit idle noise. Substituting the error
parameters of each strategy into Eq.~\eqref{eq:bell-logical-error}
yields the curves in Fig.~\ref{fig:crossover}.
Strategy~2 eliminates the idle penalty on local data qubits between
syndrome-measurement rounds
($\tilde{p}_{\mathrm{local}}^{\mathrm{S2}}=p_{\mathrm{phys}}$),
but stores the earliest Bell pair for $\mathcal{O}(d_s)$
times longer than in Strategy~1.
From Eqs.~\eqref{eq:fidelity_S1_app} and~\eqref{eq:fidelity_S2},
\begin{equation}
\label{eq:fidelity_ratio}
\frac{F_{\mathrm{stored}}^{\mathrm{S2}}}{F_{\mathrm{stored}}^{\mathrm{S1}}}
= \exp\!\Bigl(-\frac{(d_s-1)\,n^{\mathrm{round}}}
                     {\eta_{\mathrm{link}}}\Bigr),
\end{equation}
so Strategy~2's Bell-pair fidelity degrades exponentially with an extra factor
of $d_s$ compared to Strategy~1. By contrast,
Strategy~1's idle penalty $p_{\mathrm{idle}}$
(Eq.~\eqref{eq:idle_error}) has exponent
$n^{\mathrm{round}}/(\mu\,\eta_{\mathrm{link}})$, linear in $d_s$
via $n^{\mathrm{round}}=a d_s - c$. Both exponents are evaluated
at the same $d_s$, prior to the self-consistency iteration of
Algorithm~\ref{alg:self_consistent} that adjusts $d_s$ upward in
response to the accumulated decoherence.

Since both $\tilde{p}_{\mathrm{Bell}}$ and $\tilde{p}_{\mathrm{local}}$ enter the
logical error rate (Eq.~\eqref{eq:bell-logical-error}) raised to
an exponent that grows with~$d_s$, the exponential fidelity gap
between strategies is further amplified at larger code distance,
whereas the moderate increase in $\tilde{p}_{\mathrm{local}}$ from idle
noise has a comparatively weak effect.
Full numerical evaluation confirms that Strategy~1 dominates across the region where both strategies are feasible (Fig.~\ref{fig:crossover});
at sufficiently large $\eta_{\mathrm{link}}$ both error
contributions vanish and the two strategies converge. We therefore adopt Strategy~1 as the default scheduling protocol
throughout this work.

\paragraph{Numerical simulation}

We simulate a lattice-surgery merge-and-split cycle between two
distance-$d_s$ rotated surface-code patches for $d_s \in \{3, 5, 7\}$.
The circuit comprises one initialization round, $d_s$
syndrome-extraction rounds with the seam active (the last of which
is the split round), one post-split round, and final data-qubit
readout.  All local gate, measurement, and reset noise follows
Table~\ref{tab:noise_parameters} with $p_{\mathrm{phys}} = 10^{-3}$.

Each seam CNOT gate is realized by teleporting the gate through a
shared Bell pair, consumed in the zigzag order shown in
Fig.~\ref{fig:remote_surgery}(c).
The resulting effective noise channel on
$(\mathrm{ctrl},\,\mathrm{target})$ combines Bell-pair
depolarization at rate~$p_{\mathrm{Bell}}$ with five local
operations each at rate~$p_{\mathrm{phys}}$, and is dominated by the three nontrivial coset representatives
$I\!\otimes\!X$, $Z\!\otimes\!I$, and $Z\!\otimes\!X$.
Non-seam CNOT gates carry only standard depolarizing noise
(Table~\ref{tab:noise_parameters}).

For Strategy~1 the idle time between syndrome rounds adds
single-qubit depolarizing noise at rate~$p_{\mathrm{idle}}$
(Eq.~\eqref{eq:idle_error}) on every data qubit
before each of the $d_s-1$ post-merge rounds. Results for $\mu=10$ are presented in Fig.~\ref{fig:crossover}; this larger value suppresses Strategy~1's idle penalty and isolates the Bell-pair fidelity gap.

At fixed initial fidelity~$F_0$ and local error
rate~$p_{\mathrm{phys}}$, the noise budget of each strategy is
primarily described by the dimensionless parameter
$\eta_{\mathrm{link}} \equiv \lambda\,\tau_{\mathrm{coh}}$, the expected
number of Bell pairs generated per memory coherence time.
Physically, a batch of $n^{\mathrm{round}}$ pairs takes time
$n^{\mathrm{round}}/\lambda$ to collect; dividing by
$\tau_{\mathrm{coh}}$ gives the fractional coherence consumed per
round, $n^{\mathrm{round}}/\eta_{\mathrm{link}}$, which appears in the fidelity decay (Eq.~\eqref{eq:fidelity_S1_app}) and, for Strategy~1,
the idle error (Eq.~\eqref{eq:idle_error}).
A larger $\eta_{\mathrm{link}}$ therefore means a faster source
relative to decoherence, yielding lower~$p_{\mathrm{Bell}}$ for
both strategies and lower~$p_{\mathrm{idle}}$ for Strategy~1.
In the figures, $\eta_{\mathrm{link}}$ serves as the horizontal
axis; for each value, $p_{\mathrm{Bell}}$ and (where applicable)
$p_{\mathrm{idle}}$ are computed from Eqs.~\eqref{eq:fidelity_S1_app},
\eqref{eq:idle_error}, and \eqref{eq:fidelity_S2}
with $F_0 \in \{0.99,\, 0.95\}$ and
$n^{\mathrm{round}} = 2d_s - 1$ (Eq.~\eqref{eq:round_pairs}),
then injected into the circuit as described above.

Both $Z_L\!\otimes\!Z_L$ and $X_L\!\otimes\!X_L$ measurement
circuits are simulated.
$\eta_{\mathrm{link}}$ is swept over
16 logarithmically spaced points in $[10^{3},\,5\times10^{5}]$.
Circuits are built in Stim~\cite{stim} and decoded with
minimum-weight perfect matching
(PyMatching~\cite{pymatching}) using $10^{5}$ total shots
per $(d_s,\,F_0,\,\eta_{\mathrm{link}})$ point,
split equally between $Z_L\!\otimes\!Z_L$ and $X_L\!\otimes\!X_L$.
The basis-averaged logical error rate is
$p_L = \tfrac{1}{2}(p_{L,ZZ}+p_{L,XX})$,
where each $p_{L,\mathcal{B}}$ is the fraction of shots
with at least one logical flip among the three observables
$\{O_L^{\mathrm{left}},\;O_L^{\mathrm{right}},\; O_L^{\mathrm{left}}\!\otimes\!O_L^{\mathrm{right}}\}$ in basis~$\mathcal{B}$.

Figure~\ref{fig:crossover} shows that S1 outperforms S2 at
low-to-moderate $\eta_{\mathrm{link}}$, where the shorter per-round
storage time keeps Bell-pair fidelity significantly higher, despite
the additional idle depolarization
$p_{\mathrm{idle}}(\eta_{\mathrm{link}})$ on data qubits between
rounds; as $\eta_{\mathrm{link}} \to \infty$ both Bell-pair noise
and idle noise vanish, so the two strategies converge.

\begin{table}[ht]
\centering
\caption{Time-scale hierarchy in distributed QEC. Circuit durations are
quoted as circuit depth (the number of parallel layers), in units of the
two-qubit-gate time $t_{\mathrm{gate}}$, not as raw gate counts.}
\label{tab:timing}
\begin{tabular}{lll}
\hline\hline
Time scale & Symbol & Typical value \\
\hline
Two-qubit gate          & $t_{\mathrm{gate}}$    & $1$ (reference) \\
Measurement             & $t_{\mathrm{meas}}$    & $\sim t_{\mathrm{gate}}$ \\
Syndrome extraction     & $\tau_{\mathrm{SE}}$    & $\sim 5\,t_{\mathrm{gate}}$ \\
Distillation circuit    & $\tau_{\mathrm{D}}$     & $3$--$18\,t_{\mathrm{gate}}$ \\
Memory coherence        & $\tau_{\mathrm{coh}}$   & $\gg \tau_{\mathrm{SE}}$ \\
Data-qubit depolarising & $\tau_{\mathrm{dep}}$   & $\mu\,\tau_{\mathrm{coh}}$ \\
\hline\hline
\end{tabular}
\end{table}

\section*{Data availability}
The numerical data underlying all figures are available at
\url{https://github.com/sitong1011/remotels} upon request and
will be made publicly available upon publication.

\section*{Code availability}
Simulation code for the lattice-surgery scheduling comparison
(Sec.~\ref{app:scheduling_comparison}), including Stim circuit
implementations and PyMatching decoding scripts, is available at
\url{https://github.com/sitong1011/remotels} upon request and
will be made publicly available upon publication.

\begin{acknowledgments}
\label{sec:acknowledgements}
The authors are grateful to Frank Mueller, Pedro Lopes, and Abhinav Anand for helpful discussions. Sitong Liu acknowledges support from the National Science Foundation STAQ project under Grant No. PHY-2325080. This research used resources of the National Energy Research Scientific Computing Center, a DOE Office of Science User Facility supported by the Office of Science of the U.S. Department of Energy under Contract No. DE-AC02-05CH11231 using NERSC award ASCR-ERCAP0037552. This work was also supported in part by the U.S. Department of Energy, Office of Science, under Award No. DE-SCL0000039 to Lawrence Berkeley National Laboratory (PI: Erhan Saglamyurek). John Stack acknowledges partial support from NSF OMA-2120757, PHY-2325080 and DOE DE-SC0025384. Numerical simulations used the \texttt{qevo} package
(\url{https://github.com/Krastanov/qevo}) for optimized entanglement
distillation protocols, and Stim~\cite{stim} and PyMatching\cite{pymatching} for
circuit-level noise simulation and decoding. This work was initiated and primarily led by the first author, based on research conducted at Lawrence Berkeley National Laboratory in the summer of 2025.

\end{acknowledgments}
% \section*{Author Contributions}
% The project was conceived and planned by S.L., E.S., and K.K., and supervised by E.S. and K.K. with support from I.M. and K.R.B. Theoretical and numerical analyses were performed by S.L., with contributions from J.S., under the guidance of E.S., K.K., R.V.B. and K.R.B. The hardware implementation was analyzed with input from K.R.B., K.S., and E.S. The first draft was written by S.L. All authors contributed to revising and completing the manuscript.

\section*{Competing Interests}
Author KRB is a shareholder of IonQ and an advisor for Logiqal. All other authors declare no competing interests.

%\nocite{*}

\bibliography{apssamp}% Produces the bibliography via BibTeX.

\clearpage

\section*{Supplementary Information}
\appendix
\subsection{Distillation execution models}
\label{app:distillation_execution}

\paragraph{Serial execution with restarts}

If distillation attempts are executed sequentially until success, the expected number of attempts required to obtain one purified pair is $1/p_{\mathrm{succ}}$. The expected raw Bell-pair cost per syndrome extraction round is therefore

\begin{equation}
\mathbb{E}[N_{\mathrm{raw}}^{(\mathrm{serial})}]
= \frac{n_{\mathrm{pairs}}}{p_{\mathrm{succ}}} \cdot n^{\mathrm{round}}(p_{\mathrm{eff}}),
\label{eq:serial_restart}
\end{equation}
where each distillation attempt consumes $n_{\mathrm{pairs}}$ raw pairs with success probability $p_{\mathrm{succ}}$, and each syndrome extraction round requires $n^{\mathrm{round}}(p_{\mathrm{eff}})$ distilled pairs at error rate $p_{\mathrm{eff}}$.

In the most conservative restart structure, any measurement failure contaminates the target Bell pair and forces a complete restart of the protocol, so that all raw Bell pairs used in the failed attempt are discarded. This \emph{full-restart} behavior is exemplified by double-selection purification~\cite{Fujii2008EntanglementPW}, where $n_{\mathrm{pairs}}=3$ and a failure at either selection
step requires restarting the entire circuit.

More generally, some distillation circuits admit selective-retry structures, in which only the failed subcircuits and their dependent operations must be re-executed. Such protocols reduce the expected raw Bell-pair consumption relative to the full-restart case at fixed $p_{\mathrm{succ}}$. In either case, Eq.~(\ref{eq:serial_restart}) provides the baseline against which parallel execution is compared.

\paragraph{Parallel execution}

To achieve $\geq 99\%$ success probability per syndrome extraction round, we execute $k$ independent distillation attempts in parallel,
with $k$ chosen~\cite{leone2024resourceoverheadsattainablerates}
to satisfy $1-(1-p_{\mathrm{succ}})^k \geq 0.99$, giving

\begin{equation}
k = \left\lceil \frac{\log(0.01)}{\log(1 - p_{\mathrm{succ}})} \right\rceil ,
\label{eq:multiplexing_factor}
\end{equation}

The total raw Bell-pair cost per syndrome extraction round is then

\begin{equation}
C^{\mathrm{round}} = n^{\mathrm{round}} \times n_{\mathrm{pairs}} \times k,
\label{eq:cost_parallel}
\end{equation}
where $n^{\mathrm{round}} = ad_s^{*}(p_{\mathrm{eff}}) - c$ is the number of purified pairs required per syndrome extraction round, $n_{\mathrm{pairs}}$ is the raw input consumption per distillation attempt, and $k$ is the multiplexing factor that ensures $\geq 99\%$ success probability per round.

If successfully distilled Bell pairs can be buffered across rounds, the long-term average consumption converges to the serial rate: each batch of $k$ parallel attempts yields an expected $k \cdot p_{\mathrm{succ}}$ successes while consuming $k \cdot n_{\mathrm{pairs}}$ raw pairs, giving a cost per success of $n_{\mathrm{pairs}}/p_{\mathrm{succ}}$, identical to Eq.~(\ref{eq:serial_restart}).

\subsection{\label{app:phyQ}Physical qubit budget}

\begin{figure*}[ht]
  \centering
{\includegraphics[width=0.49\textwidth]{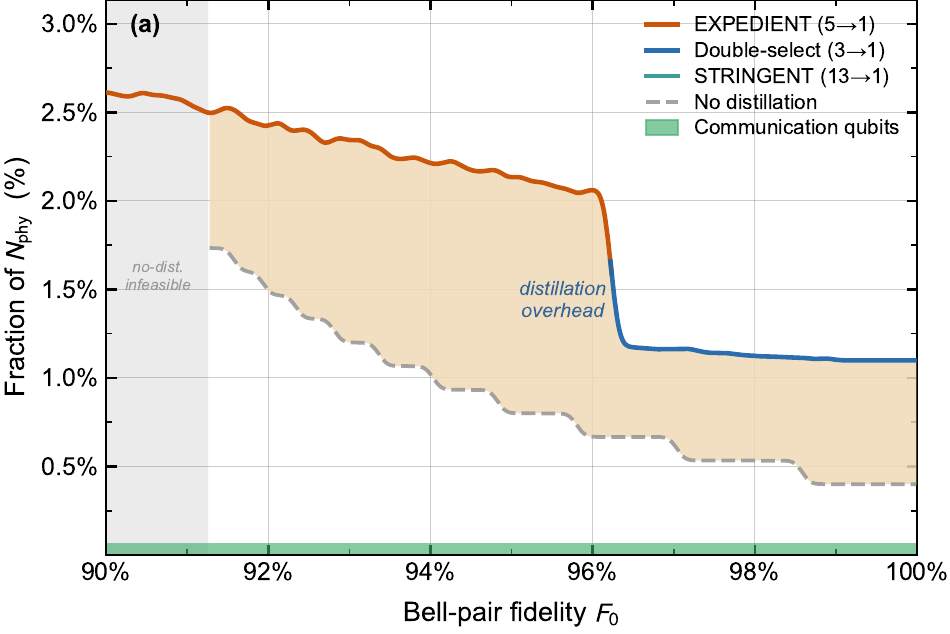}
    \label{fig}}
\hfill
\includegraphics[width=0.49\textwidth]{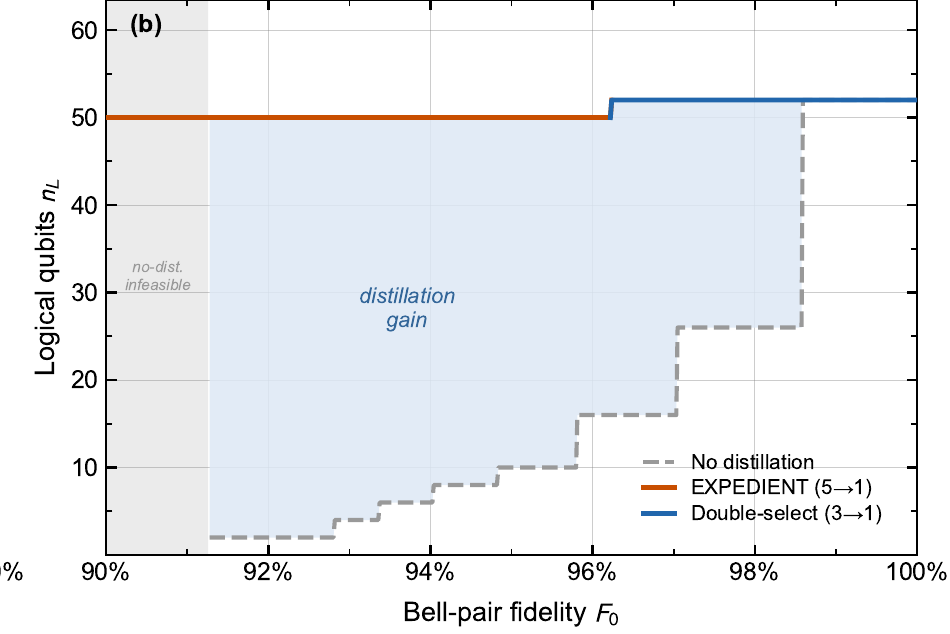}%
\caption{%
  \textbf{Physical qubit allocation per module vs.\ Bell-pair fidelity
  $\bm{F_0}$ ($\bm{N_{\mathrm{phy}}=3000}$,
  $\bm{p_L\leq10^{-3}}$).}
  (a)~Communication and memory physical qubit consumption per
  module.
  The green band shows the fixed communication overhead $N_{\mathrm{comm}}$
  (Eq. ~\eqref{eq:interfaces}), set to the minimum communication-qubit count needed to support $I = 2$ optical interfaces per module. Solid colored lines show the optimal distillation envelope (protocol with highest $n_L$ at each $F_0$); the orange band indicates the extra memory $N_{\mathrm{mem}}$ incurred by distillation relative to no distillation (gray dashed line; Eqs. \eqref{eq:memo_dist} and \eqref{eq:memo_raw}).
  (b)~Logical qubit capacity $n_L$ (Eq.~\eqref{eq:nL}),
  assuming a two-column patch layout per module that accounts for
  inter-patch lattice-surgery overhead. Solid colored lines show the optimal distillation envelope; the blue band indicates the distillation gain $\Delta n_L>0$ relative to no distillation (gray dashed line). The gray region marks the no-distillation infeasible zone ($n_L<2$). Noise parameters are listed in Table~\ref{tab:noise_parameters};
  surface-code distances $d_s^*$ are computed following
  Sec.~\ref{subsec: fitted}.%
}

  \label{fig:phy_budget}
\end{figure*}

Here we assume that before any logical encoding begins, each module in a distributed architecture incurs a fixed
physical-qubit overhead. Memory qubits are needed to buffer Bell pairs during stochastic generation.
Distillation reduces code distance at the cost of higher per-pair
consumption. In contrast, direct consumption avoids that consumption overhead but
requires a larger code distance to tolerate raw noise. Consequently, which strategy
yields more logical qubits from the same $N_{\mathrm{phy}}$ depends
on the operating point. 

\paragraph{\label{subsub:comm}Communication and memory qubits}

\textit{Communication qubits.}\;
Each of the $I$ optical interfaces requires dedicated qubits to mediate
photon emission and heralded entanglement. The minimum number of qubits used to generate entanglement is therefore
$N_{\mathrm{comm}} = I$~\cite{Singh_2025, Main_2025}. When the attempt
rate exceeds the inverse reset time, time-division multiplexing
is required to maintain throughput, giving
\begin{equation}
\label{eq:interfaces}
N_{\mathrm{comm}} = I \cdot \max\!\left(1,\;
\left\lceil \tau_{\mathrm{reset}} \cdot r_{\mathrm{attempt}}
\right\rceil\right).
\end{equation}
This count is set entirely by generation-layer hardware and is
independent of code distance or distillation strategy.

\textit{Memory qubits.}\;
Each heralded pair occupies a memory qubit until it is consumed. We
assume the conservative minimum occupancy, setting aside decay effects
analyzed in Sec.~\ref{subsec:decay}.

Without distillation, the module stores one raw pair per seam qubit
per syndrome round:
\begin{equation}
N_{\mathrm{mem}}^{(\mathrm{raw})}
= n^{\mathrm{round}}(p_{\mathrm{raw}})
\label{eq:memo_raw}
\end{equation}
with distillation, each of the
$n^{\mathrm{round}}(p_{\mathrm{eff}})$ purified pairs needed per round
is produced by $k$ parallel distillation circuits
(Eq.~\eqref{eq:multiplexing_factor}), each holding
$n_{\mathrm{pairs}}$ raw inputs simultaneously:
\begin{equation}
N_{\mathrm{mem}}^{(\mathrm{dist})}
= n^{\mathrm{round}}(p_{\mathrm{eff}})
  \times k \times n_{\mathrm{pairs}}.
\label{eq:memo_dist}
\end{equation}
Whether $N_{\mathrm{mem}}^{(\mathrm{dist})}$ exceeds
$N_{\mathrm{mem}}^{(\mathrm{raw})}$ depends on whether the distance
reduction from distillation compensates for the
$k \times n_{\mathrm{pairs}}$ multiplier.

\paragraph{\label{subsub:logical}Logical qubit capacity}

A distance-$d$ rotated surface code requires
$2d^{2}-1$ physical qubits per patch
($d^2$ data, $d^2-1$ ancilla); ancilla-reuse
schemes~\cite{Chatterjee2024QuantumPD} reduce this to
$\approx 1.5\,d^2$, though we adopt the conservative count.
Lattice surgery between adjacent patches requires
$d$~additional physical qubits along each shared boundary.
For simplicity we consider a two-column patch grid,
in which $n_L/2$ rows share $\tfrac{3}{2}n_L - 2$
internal boundaries.
After subtracting overhead, a module hosts
\begin{equation}
\label{eq:nL}
  n_L \;=\; 2\left\lfloor
    \frac{N_{\mathrm{phy}} - N_{\mathrm{comm}}
          - N_{\mathrm{mem}} + 2d_s}
         {4(d_s)^{2} + 3\,d_s - 2}
  \right\rfloor
\end{equation}
logical qubits. The denominator grows as $(d_s)^2$, so
lowering $p_L^{\mathrm{target}}$ quadratically compresses logical
capacity even before overhead is accounted for.

Distillation reduces $d_s$; to leading order the capacity gain is
$\Delta n_L \approx n_L^{\mathrm{raw}}(\rho^{-2}-1)$
where $\rho = d_s^{\mathrm{dist}}/d_s^{\mathrm{raw}}$
(Eq.~\eqref{eq:distance_ratio}), though the actual gain is
moderated by the increased memory overhead $N_{\mathrm{mem}}$.

\paragraph{\label{subsub:budget}Budget constraint}

The total physical qubit count per module satisfies
\begin{equation}
N_{\mathrm{phy}}
= \underbrace{n_L\bigl(2(d_s)^2 - 1\bigr)
  + \bigl(\tfrac{3}{2}n_L - 2\bigr)\,d_s}_{\text{logical grid}}
\;+\; N_{\mathrm{comm}}
\;+\; N_{\mathrm{mem}}.
\label{eq:total_budget}
\end{equation}
Distillation shrinks the grid term through a smaller $d_s$
but increases $N_{\mathrm{mem}}$ through the
$k \times n_{\mathrm{pairs}}$ multiplier;
the optimal strategy at each $F_0$ is the one that maximises $n_L$.
Figure~\ref{fig:phy_budget} shows how the allocation shifts
across operating regimes.

\subsection{Transversal methods}
\label{app:transversal}

An alternative approach for DQC involves distributed transversal operations, as detailed in \cite{stack2025assessingteleportationlogicalqubits}. A comprehensive comparison is outside the scope of this work and is complicated by the differences in compilation methods between lattice surgery and transversal approaches. However, transversal methods generally require $n$ Bell pairs to be available for simultaneous consumption for an $[[n,k,d]]$ code \cite{stack2025assessingteleportationlogicalqubits}. This contrasts with the lattice surgery methods discussed in this paper, which typically require fewer simultaneous pairs, although the cumulative Bell pair count across all rounds is likely higher. That being said, transversal operations on dense qLDPC codes enable parallel non-local CNOTs between or teleportation of multiple logical qubits, significantly lowering the number of Bell pairs used per logical operation \cite{stack2025assessingteleportationlogicalqubits}. 

%
% ****** End of file apssamp.tex ******

\end{document}